\documentstyle[12pt,aaspp4]{article}

\newcommand {\hii}{H {\small II}~}
\newcommand {\ha}{H$\alpha$~}
\newcommand {\flux}{ergs^{-1}cm{-2}~}
\newcommand {\lum}{ergs^{-1}}
\newcommand{\kms}{\ifmmode {\rm \, km \, s^{-1}} \else $\rm \,km, s^{-1}$\fi}

\begin{document}

\title{H$\alpha$ Imaging of Early-Type(Sa-Sab)
 Spiral Galaxies I\footnote{Based on observations obtained with the 
1.5-meter telescope at CTIO and the 3.5-meter telescope at Apache 
Point Observatory (APO). The APO 3.5m telescope is owned and operated by the 
Astrophysical Research Consortium.}}

\author{Salman Hameed\altaffilmark{2}, Nick Devereux\altaffilmark{2}}

\affil{Astronomy Department, New Mexico State University, Las Cruces, NM 88003}

\altaffiltext{2}{Visiting Astronomer, Cerro Tololo Inter-American Observatory.
CTIO is operated by the Association of Universities for Research in Astronomy, Inc. (AURA)
under cooperative agreement with the National Science Foundation.}

\authoremail{shameed@nmsu.edu, devereux@nmsu.edu}

\begin{abstract}

\ha and continuum images are presented for 27 nearby early-type(Sa-Sab) 
spiral galaxies. Contrary to popular perception, the images reveal
copious massive star formation in some of these galaxies. A
determination of the \ha morphology and a measure of the \ha
luminosity suggests that early-type spirals can be classified into two
broad categories based on the luminosity of largest \hii region in the
disk. The first category includes galaxies for which the individual
\hii regions have $L_{H\alpha} < 10^{39} erg s^{-1}$. Most of the
category 1 galaxies appear to be morphologically undisturbed, but show
a wide diversity in nuclear \ha properties. The second category
includes galaxies which have at least one \hii region in the disk with
$L_{H\alpha} \ge 10^{39} erg s^{-1}$. All category 2 galaxies show
either prominent dust lanes or other morphological peculiarities such
as tidal tails which suggests that  the anomalously luminous \hii regions in category 2
galaxies may have formed as a result of a recent interaction. The
observations, which are part of an on-going \ha survey, reveal
early-type spirals to be a heterogeneous class of galaxies that are
evolving in the current epoch. 

We have also identified some systematic differences
between the classifications of spiral galaxies in the 
Second General Catalog (RC2) and the Revised Shapley-Ames 
Catalog (RSA) which may be traced to subtle variations in the application
of the criteria used for classifying spiral galaxies. An examination of
earlier studies suggests that perceptions concerning the Hubble type
dependence of star formation rates among spiral galaxies 
depends on the choice of catalog.

\end{abstract}


\section{Introduction}

An outstanding problem in extra-galactic astronomy is 
understanding the parameters that determine the structure and 
evolution of galaxies. A first step towards understanding the physical 
properties of galaxies is classification based on morphology.

Spiral galaxies were first classified by Hubble according to the size
of the bulge, the tightness with which the spiral arms are wound, and
the resolution of individual \hii regions (\markcite{Hubble1936}Hubble
1936;
\markcite{devauc1959} de Vaucouleurs 1959). 
Several observational studies have elucidated differences among
galaxies along the Hubble sequence (for a review see
\markcite{Roberts1994} Roberts \& Haynes 1994;
\markcite{Kennicutt1998} Kennicutt 1998). 
Observations plus modeling of broad band colors of galaxies reveal
that the bulge dominated Sa galaxies are red compared to the
disk dominated Sc galaxies, suggesting an older population
\markcite{Larson1974}(Larson \& Tinsley 1974).  Similarly,
\markcite{Roberts1969}Roberts(1969) measured the atomic gas content of
75 spiral galaxies and found that the ratio, M(HI)/L(B), decreases
systematically from the late-type spirals to the early-type Sa
galaxies, suggesting early types to be deficient in hydrogen gas,
compared to the late types.

The star formation history of galaxies can be traced using the integrated 
\ha equivalent widths, where the
\ha emission line flux is normalized by the past star formation
rate through the red continuum. \markcite{KK1983}Kennicutt \& Kent
(1983) measured the
\ha equivalent widths for $\sim 200$ spiral galaxies and demonstrated 
that the \ha equivalent widths decrease systematically from late-type
spirals to early-type spiral galaxies, suggesting early-types to be
deficient in massive young stars compared to the late types.

There are other observational results, however, which suggest that 
early-type spirals are not so quiescent. A recent Hubble Space 
Telescope(HST) study of the 
bulges of 75 spiral galaxies (\markcite{Carollo1998}Carollo {\it et al.} 1998) reveals a 
wide variety of activity, including star formation, hidden underneath 
the bulges of early-type spirals. \markcite{YK1989}Young \& Knezek (1989) 
have showed that the dominant phase of the interstellar medium 
in Sa-Sab types is molecular, not atomic and that the molecular 
fraction is much higher in the early-types compared to the later types.
The result, however,
has been recently challenged by \markcite{Casoli1998}Casoli {\it et al.} (1998), who find that
molecular gas comprises only about one third to one fourth of
the total gas content of spirals of types Sa through Sc.

A recent analysis of the Infrared Astronomical Satellite (IRAS) database 
by \markcite{DH1997}Devereux \& Hameed (1997)
suggests that the global massive star formation rates, as determined by 
60 micron luminosity functions, are comparable in early and 
late-type spirals. Similarly, far infrared to blue luminosity ratios of a large 
sample of nearby spiral galaxies do not show any morphological dependence 
(\markcite{Tomita1996}Tomita {\it et al.} 1996; \markcite{DH1997}Devereux \& Hameed 1997). 
Evidently, the IRAS data has revealed  a previously 
unsuspected population of early-type spirals with high massive star formation 
rates. 

The IRAS results, do not support previous claims,
based on \ha equivalent widths, that massive star formation rates increase 
along the Hubble sequence from Sa to Sc(\markcite{Kennicutt1983}Kennicutt 1983; 
\markcite{Kennicutt1994}Kennicutt {\it et al.} 1994). 
Part of the problem is that the sample of early-type spirals selected by 
\markcite{Kennicutt1983}Kennicutt (1983)
and \markcite{Kennicutt1994}Kennicutt (1994) is small in number and is biased towards galaxies 
with low values of L(FIR)/L(B) (\markcite{DH1997}Devereux \& Hameed 1997; 
\markcite{Usui1998}Usui {\it et al.} 1998). 
We are, therefore, conducting 
an \ha imaging survey of {\it all known} nearby early-type spiral galaxies in order to better 
understand the differences between the IRAS results and those of existing \ha studies. 

High resolution \ha images of nearby galaxies provide important  
information about the morphology and luminosity of the ionized 
hydrogen gas. Surprisingly few \ha images of early-type spirals 
exist in the published literature. The general notion that early-type 
spirals do not have significant massive star formation is, at least 
partially, responsible for the dearth of \ha observations 
(\markcite{Young1996}Young {\it et al.} 1996).
The continuum morphology of
early-type spirals is dominated by an 'inert' stellar bulge which 
can hide star forming complexes that lie underneath them. CCD imaging  
allows the hidden \hii regions to be revealed by subtracting the 
overwhelming continuum light.

Despite the dearth of \ha images,
our appreciation of the heterogeneous nature of early-type spirals has 
evolved considerably in the past quarter century. 
\markcite{vandenbergh1976}Van den Bergh(1976) and 
\markcite{Kormendy1977}Kormendy(1977) found no \hii regions in 
NGC 4594 and NGC 2841, leading them to speculate that the 
IMF in early-type spirals may be biased against the formation of
massive stars. Later studies, however,
showed that \hii regions are indeed present in these particular galaxies 
(\markcite{Schweizer1978}Schweizer 1978; \markcite{HK1983}Hodge \& Kennicutt 1983; 
\markcite{Kennicutt1988}Kennicutt 1988).
A detailed study of \hii regions in the disks of seven Sa galaxies 
by \markcite{Caldwell1991}Caldwell {\it et al.} (1991) found that \hii regions are quite abundant in 
the disks but  are significantly smaller than those in late-type spirals.
Specifically, \markcite{Caldwell1991}Caldwell {\it et al.} found that there are no \hii regions 
in the disks of early-type spirals with luminosities $ >10^{39} erg s^{-1}$.

The purpose of the present paper is to report that \hii regions are not only 
abundant in early-type spirals but some contain giant \hii 
regions that are comparable in size and luminosity to giant \hii regions 
seen in late-type spirals. Our findings support recent \ha observations
by \markcite{Young1996}Young {\it et al.} (1996) and 
\markcite{Usui1998}Usui {\it et al.} (1998), who have also identified
numerous early-type spirals with star formation rates comparable to
the most prolifically star forming late-type spirals.  

In order to quantify the diverse star forming capabilities of early-type 
spirals, we are conducting a systematic program of \ha imaging. 
The results are  presented here for twenty seven galaxies imaged to date. 
The sample is described in section 2 and the observations are described 
in section 3. The results are presented in 
section 4, followed by discussion in section 5.

\section{The Sample}

An \ha imaging survey is being conducted to investigate the star
forming capabilities of nearby early-type (Sa-Sab) spiral
galaxies. The target galaxies have been selected from the Nearby
Galaxy Catalog (NBG)(\markcite{Tully1988}Tully 1988) which is the largest
complete compilation of bright galaxies with velocity less than $3000 km/s$,
corresponding to a distance of 40 Mpc
($H_{0}=75 kms^{-1}/Mpc$). The target galaxies are listed in Table 1
with some useful observables. Distances in the NBG  catalog are based
on a Virgo-centric in-fall model (\markcite{Tully1988}Tully 1988).
The NBG catalog also lists morphological types for each galaxy 

The complete sample includes
all (57) bright, m(B) $\le$ 12.1 magnitude,  non-interacting early-type (Sa-Sab) spirals 
known within 40 Mpc. For the purposes of the present work, ``interacting'' galaxies
are defined as those which have cataloged companions 
within 6\arcmin  of each other.

The goal of the survey is to image the complete sample of nearby 
early-type spirals. Imaging the largest complete
sample will circumvent incompleteness corrections and 
minimize statistical errors in quantifying the incidence of nuclear starbursts, 
nuclear emission line spirals, nuclear point sources, and other morphological 
peculiarities in nearby early-type spirals. So far we have imaged twenty-one
galaxies, the results of which are presented in  this paper. 

As part of a complimentary study we are also obtaining \ha images of twenty-one 
far-infrared luminous early-type spirals,  
identified by  \markcite{DH1997} Devereux and Hameed (1997). Fifteen of these galaxies 
have m(B) $\le$ 12.1 and, hence, are already included in the complete sample described above. 
The remaining six galaxies have been imaged and they are included 
in this paper also.

\section{Observations}

Observations of northern hemisphere galaxies were obtained 
with the Astrophysical Research Consortium (ARC) 3.5m telescope at 
Apache Point Observatory (APO) in New Mexico. Southern hemisphere 
galaxies were observed with the
1.5m telescope at Cerro Tololo Inter-American Observatory (CTIO) 
in Chile. 

Six northern hemisphere early-type spiral galaxies were observed using
the Double Imaging Spectrograph (DIS) at APO between August 1996 and
January 1997. The Texas Instruments CCD chip has a pixel scale of
$0.61\arcsec pixel^{-1}$ and a $4.2\arcmin$ field of view.  Two
red-shifted narrow band H$\alpha$ + [NII] (6570\AA \& 6610\AA,
$\Delta\lambda$ = 72\AA) filters and a line free red continuum
(6450\AA,$ \Delta\lambda$ = 120\AA) filter were used to obtain the
line and continuum images, respectively. Three exposures were obtained
through each of the line and continuum filters. Details of the
observations are summarized in Table 2.

Twenty-one southern hemisphere galaxies were observed using the 
Cassegrian Focus CCD Imager (CFCCD) on the CTIO 1.5m telescope. 
CFCCD uses a $2048 \times 2048$ Tektronics chip 
and has a pixel scale of $0.43\arcsec pixel^{-1}$ at f/7.5, yielding 
a field of view of $14.7' \times 14.7'$. All of the 
galaxies were imaged using the narrow band H$\alpha$ + [NII] filter at 
6606\AA ($\Delta\lambda$ = 75\AA) with the exception of NGC 5728 for which 
we used the H$\alpha$ + [NII] filter at 6649\AA ($\Delta\lambda$ = 76\AA). 
The narrow band line free red continuum filter for all the galaxies was 
centered at 6477\AA ($\Delta\lambda$ = 75\AA). Three exposures of 900 seconds 
 were obtained through each of the line and the continuum 
filters for all the southern hemisphere galaxies (see Table 2). 

\subsection{Data Reduction}

The Image Reduction Analysis Facility (IRAF) software package was used to process the images. 
The images were bias subtracted, and then flat fielded using twilight flats taken on 
the same night as the galaxy data. Sky subtraction 
was achieved by measuring the sky level around the galaxy 
and then subtracting it from the images. Images were then registered and median combined to 
improve the S/N ratio and eliminate cosmic rays. 

The continuum image was then scaled to the line plus continuum image by 
measuring the integrated fluxes of several ($\ge 10$) stars common 
to both images. The final 
H$\alpha$ image was obtained by subtracting the scaled continuum image
from the H$\alpha$+continuum image to remove foreground stars 
and the galaxy continuum. 

The H$\alpha$ images were flux calibrated using observations of the
standard stars G1912B2B and BD $+28\arcdeg 4211$
(\markcite{Massey1988}Massey {\it et al.} 1988) for the northern
hemisphere galaxies and LTT 3218, LTT 1020, and LTT 7987
(\markcite{Hamuy1994}Hamuy {\it et al.} 1994)  for the southern hemisphere
galaxies (see Table 2). The observing conditions were photometric,
with $<10 \%$ variations in the standard star fluxes, for all of the
imaging observations.

There are several factors that contribute to the uncertainties
in the determination of \ha flux measurements. 
Large systematic errors may be introduced by the continuum 
subtraction procedure. Small random errors are introduced by read noise 
in the detector and photon noise but these are negligible compared to the 
uncertainty introduced by the continuum subtraction procedure. 

There are two factors that contribute to the uncertainty
in determining the continuum level. First, one assumes that the 
foreground stars used for estimating continuum level
have the same intrinsic color as the galaxy. Second, one must assume that the 
galaxy has the same color everywhere. Both assumptions are
unlikely to be correct. Unfortunately the uncertainty in the 
\ha flux depends sensitively and non-linearly on the 
continuum level subtracted. It has been determined empirically that 
$2\%-4\%$ errors in the continuum level correspond to 
$10\%-50\%$ errors in the \ha flux, depending on the relative 
contribution of the continuum
 light.

The \ha fluxes are presented in Table 3 along with the calculated 
\ha luminosities. The \ha fluxes include contributions from the
satellite [NII] lines at $\lambda\lambda$ 6548,6584 and have  
not been corrected for Galactic or internal extinction.\markcite{KK1983}
Kennicutt and Kent (1983) suggest that the average extinction in
their sample is typically $\sim$1 mag. The \ha extinction, however, is
expected to be more for galaxies that have high inclinations or those
that harbor nuclear starbursts. We have not applied extinction corrections
due to the uncertainties involved in determining its true value.
Consequently the \ha fluxes and luminosities
presented in Table 3 are lower limits to the intrinsic \ha fluxes and
luminosities.

\subsection{Comparison with Previous Measurements}

Seven of the galaxies presented in this paper already have published
\ha fluxes. Figure 1 compares our measured values in the same apertures as
those in the literature.
The agreement between the measurements is, in general, good.  NGC 1433
and NGC 7552, in particular, provide a good test as both ours and the
comparison observations(\markcite{Crocker1996}Crocker {\it et al.}
1996 and \markcite{Lehnert1995}Lehnert
\& Heckman 1995 respectively) were made with the 1.5m
telescope at CTIO and the measurements are in good ($<11\%$) agreement with 
each other.  

The biggest discrepancy is with NGC 1022, where our \ha flux 
is $68\%$ higher than the value published by \markcite{KK1983}Kennicutt 
\& Kent (1983). However, our measured flux for NGC 1022 is within $8\%$ 
of the value obtained by \markcite{Usui1998}Usui {\it et al.} (1998).

\section{Results}

\subsection{Classification of Early-type Spirals}

A study of the \hii region luminosity functions in the disks of seven
Sa galaxies by \markcite{Caldwell1991}Caldwell {\it et al.}(1991)
showed that while \hii regions are abundant, there are none with \ha
luminosities $> 10^{39} erg s^{-1}$.  Our results confirm that the \ha
luminosity of \hii regions in most early-type spirals is less than
$10^{39} erg s^{-1}$. However, we also find that a significant
fraction ($15-20\%$) of early-type spirals do have at least one \hii
region in the disk with $L_{H\alpha} \ge 10^{39} erg s^{-1}$. The latter 
result is not necessarily in contrast to Caldwell's study for two reasons. 
First, Caldwell {\it et al.} had a small sample containing only
seven galaxies and could have easily missed early-type spirals 
with giant \hii regions. Second, their sample contained only Sa 
galaxies, whereas early-type spirals discussed in this paper
include both Sa and Sab types. 

Early-type spirals have been divided in two categories based on the \ha 
luminosity of the largest \hii region present in the disk. The \ha luminosity 
of the individual \hii regions in the disks of all category 1 
galaxies is less than $10^{39} erg s^{-1}$, whereas category 2 galaxies 
contain at least one \hii region in the disk with $L_{H\alpha} 
\ge 10^{39} erg s^{-1}$.

Figure 2 shows the range of global \ha luminosities for the two
categories of Sa-Sab galaxies. Perhaps not surprisingly, category
2 galaxies dominate at high, $L_{H\alpha}>1.7 \times 10^{41}\lum$,
luminosities.  A two-tailed Kolmogorov-Smirnov(KS) test indicates that
the two categories of galaxies are not derived from the same
population at a confidence level greater than $99\%$. Similarly,
Figure 3 shows the range of \ha equivalent widths, where the
continuum-free line flux is normalized by the red continuum, for the
two categories of early-type spirals. The K-S test again reveals that
the two categories of galaxies are not derived from the same
population at a confidence level greater than $99\%$.

In order to verify that the difference in the distribution of global
\ha luminosities between the two categories is not related to the 
nuclear properties, the histograms of nuclear(1 kpc)
\ha luminosities are compared in Figure 4.
The K-S test reveals that the difference between the distributions in Figure 4
 is  insufficient to reject the null 
hypothesis that the two categories are drawn from the same
population. Thus the difference between the two categories of early-type
spirals is primarily a global phenomenon unrelated to their nuclei.

In addition to the statistical tests described above, we varied the \ha luminosity 
of the largest \hii region in each galaxy by 20$\%$ to further verify the dichotomy 
in the properties of category 1 and 2 early-type spirals. Indeed, a
20 percent variation, which represents a typical uncertainty in 
\ha measurements of  individual \hii regions, has no effect on the distribution
of galaxies between the two categories. We do want to stress, however, that
the purpose of this paper is to introduce early-type spirals with 
giant \hii regions and to show that  Sa-Sab galaxies are heterogeneous 
in nature.  The sub-classification of early-type spirals into two 
categories has been done solely to investigate the conditions that 
may lead to the formation of luminous \hii regions.

The continuum and continuum-subtracted \ha images are shown in Figures
5, 6, \& 7 and the observed properties of Sa-Sab galaxies are
described in more detail below.

\begin{center}
{\it Category 1 Early-type Spirals} 
\end{center}

\hii regions in category 1 galaxies are either small or totally absent
from the spiral arms (see Figure 5). By definition, all \hii regions in the disk of
category 1 spirals have $L _{H\alpha}< 10^{39}erg s^{-1}$.  Consistent with
earlier studies (\markcite{Kennicutt1988}Kennicutt 1988;
\markcite{Caldwell1991}Caldwell {\it et al.} 1991; \markcite{Bresolin1997}Bresolin \&
Kennicutt 1997), we find that \hii regions in most early-type spirals
contain only a few massive stars, whereas \hii regions in late-type
spirals can contain hundreds or even thousands of stars.

The seventeen galaxies included in category 1 resemble what most astronomers 
identify as classical early-type spirals. Morphologically, most 
category 1 galaxies appear undisturbed in the continuum image, but 
the \ha images reveal very diverse nuclear properties. 

\noindent
{\it Extended nuclear emission line region (ENER):} There are seven
category 1 galaxies (NGC 1350, NGC 1371, NGC 1398, NGC 1433, NGC 1515,
NGC 1617, NGC 3169) in which an extended nuclear emission line
region(ENER) has been detected. The detailed morphology of the ENER
gas is difficult to discern at the distances of these galaxies, but
they appear to be similar to the nuclear emission line spirals discovered
in the centers of M81 (\markcite{Jacoby1989}Jacoby {\it et al.} 1989;
\markcite{Devereux1995}Devereux {\it et al.} 1995) and M31
(\markcite{Jacoby1985}Jacoby {\it et al.} 1985;
\markcite{Devereux1996}Devereux {\it et al.} 1996).  It is clear that
the filamentary emission line gas is quite different from the clumpy
\hii regions in the spiral arms. The ENER is most likely shock-ionized or
photo-ionized by UV radiation from bulge post asymptotic giant branch
stars (\markcite{Devereux1995}Devereux {\it et al.} 1995;
\markcite{Heckman1996}Heckman 1996).  All category 1 galaxies hosting
extended nuclear emission line regions have the lowest nuclear
\ha luminosities.

\noindent 
{\it Nuclear Point Sources:} Of the remaining category 1 galaxies,
five (NGC 2273, NGC 5188, NGC 5728, NGC 7172, NGC 7213) have an unresolved
\ha source at the nucleus. Apart from 5188, all four galaxies have
been spectroscopically identified as Seyferts(NGC 2273:
\markcite{Ho1997a}Ho {\it et al.} 1997a; NGC 5728:
\markcite{Phillips1983}Phillips {\it et al.} 1983; NGC 7172:
\markcite{Sharples1984}Sharples {\it et al.} 1984; NGC 7213:
\markcite{Filippenko1984}Filippenko \& Halpern 1984). NGC 2273, NGC 5728
and NGC 7172 have faint \hii regions in their disk, whereas NGC 7213
has a number of \hii regions in the circumnuclear region, but the disk
is almost devoid of any \hii regions. NGC 7213 also has an extended
nuclear emission line region region surrounding the point source. NGC
5188 has prominent \hii regions and its nucleus has been
spectroscopically classified as an \hii region by
\markcite{VV1986}Veron-Cetty \& Veron (1986).

\noindent
{\it Nuclear starbursts:} The nuclei in four of the remaining category
1 galaxies (NGC 3471, NGC 1482, NGC 3885, NGC 1022) are resolved and
contribute more than $50\%$ of the total \ha luminosity. NGC
1022(\markcite{Ashby1995}Ashby {\it et al.} 1995), NGC
3471(\markcite{Balzano1983}Balzano 1983) and NGC 3885
(\markcite{Lehnert1995}Lehnert \& Heckman 1995) have been
spectroscopically identified as nuclear starbursts. No spectral
confirmation is available in the literature for NGC 1482 but it has
been included as a nuclear starburst galaxy based on similarities in
morphology and
\ha luminosity to NGC 1022, NGC 3471, and NGC 3885. It is 
worthwhile to mention that all four galaxies also have very high far
infrared luminosities, $\ge 10^{10}L_{\sun}$
(\markcite{DH1997}Devereux \& Hameed 1997), comparable to the
prototypical starburst galaxy M82 (\markcite{Rieke1980}Rieke {\it et
al.} 1980).  Dust lanes are prominent in all four galaxies and there
are virtually no \hii regions in their disks.

NGC 3717 has a high inclination and could not be classified into any 
of the above sub-groups. Spectroscopically, the nucleus of NGC 3717 
has been classified as an \hii region by \markcite{VV1986}Veron-Cetty \& Veron (1986).

\begin{center}
{\it Category 2 Early-type Spirals}
\end{center}

Eight of the galaxies presented in this paper have at least one disk
\hii region with $L_{H\alpha} \ge 10^{39} erg s^{-1}$ defining  a new 
category of early-type spirals (see Figure 6). 
 
Dust lanes or tidal tails are present in all category 2 galaxies. 
Despite the fact that extinction is an important factor in these dusty 
galaxies, early-type spirals with the highest \ha luminosity belong 
exclusively to category 2(Figure 2). Additionally, 
all category 2 galaxies have far-infrared luminosities in excess of
$10^{10} L_{\sun}$ (\markcite{DH1997}Devereux \& Hameed 1997).

The discovery of a significant number of early-type spirals
with \ha luminosities comparable to those produced by  the most prolifically 
star forming late-type spirals is surprising, as early-type spirals 
are widely believed  to have massive star formation rates that are
considerably lower.
Based on the available images and far-infrared luminosities of the
entire sample, it is 
estimated that category 2 galaxies could represent $15\%-20\%$ of 
all early-type spirals in the nearby($D \le 40 Mpc$) universe, a significant 
group that can no longer be ignored as oddities. 

The continuum images of two category 2 galaxies (NGC 986, NGC 7552) reveal 
the possible existence of previously uncataloged
dwarf galaxies that appear to be  interacting with the 
larger galaxies. Redshift information is not yet available for either of 
the dwarf galaxies so the physical association cannot be confirmed. Nevertheless,
if the dwarf galaxy seen at the end of the northern 
spiral arm in NGC 7552 is indeed associated with NGC 7552, it has 
an absolute magnitude of -12.95 in R which is comparable
to some of the dwarf galaxies seen around M31 and the Milky Way.
There is no detectable \ha emission from the dwarf companion of NGC 7552. 

In the case of NGC 986, the dwarf galaxy is located 
at the end of the north-eastern spiral arm and appears to be 
in the process of being tidally disrupted. The distorted shape of
the dwarf along with the contribution of stars in the spiral arms
of NGC 986 makes it difficult to measure its magnitude with any certainty. 
Nevertheless, the presence of 
interacting dwarfs exclusively in category 2 galaxies may be related to the existence 
of giant \hii regions in these galaxies. 

Two category 2 galaxies (NGC 7582, NGC 6810) have been 
optically classified as Seyferts. The presence 
of Seyferts in both categories of early-type spirals suggest that 
the incidence of Seyfert activity  is unrelated to the phenomenon that 
is responsible for the global differences between category 1 and category 2 
galaxies.

\begin{center}
{\it Unclassified}
\end{center}

Surprisingly, NGC 660 and NGC 2146 apparently have no disk \hii
regions with $L_{H\alpha} \ge 10^{39}\lum$ but both have high far
infrared luminosity (\markcite{DH1997}Devereux \& Hameed 1997) and a morphology that
appears to be even more disturbed than any of the category 2
galaxies (see Figure 7). The apparent absence of luminous \hii regions may be a
consequence of the 13 and 9 magnitudes of extinction in the V band for
NGC 660 and NGC 2146 respectively (\markcite{Young1988}Young {\it et
al.} 1988). Nevertheless, we have not classified these two galaxies as
they cannot be unambiguously placed into either category.
  
\subsection{\hii Region Luminosity Functions}

In order to understand some of the properties of \hii regions in
the two categories of early-type spirals, we derived \ha 
luminosity functions for the \hii regions in a category 1 galaxy, 
NGC 1398, and a category 2 galaxy, NGC 7552. The two galaxies
were chosen because they are at a comparable distance (16.1 Mpc
\& 19.5 Mpc for NGC 1398 \& NGC 7552 respectively), they have 
relatively low inclinations, and they provide a good contrast 
between the two categories of early-type spirals.

\noindent{\it Measurement of \hii region \ha fluxes}

Individual \hii regions in the two galaxies were identified computationally 
using the DAOfind routine incorporated in an IDL program,'{\rm H{\small II}{\it phot}}',
developed by \markcite{Thilker1999}Thilker {\it et al.} 1999. 
The program identifies local maxima by using Gaussian kernels comparable,
and slightly larger than the effective resolution,  on the 
original image and also on several smoothed versions. 

Earlier studies of luminosity functions have mostly assumed 
a symmetrical shape for the \hii regions.(e.g. \markcite{Kennicutt1989}Kennicutt {\it et al.} 1989;
\markcite{Caldwell1991}Caldwell {\it et al.} 1991). In nature, however, \hii
regions are observed to be asymmetrical. The attribute of {\rm H{\small II}{\it
phot}} is that it provides the freedom for \hii regions to acquire any shape.
The \ha flux was measured by summing the pixel values
within each region. The local background level for each region was
determined by the mode of the 'region-hugging' annulus having a
specified physical width.  The details of the '{\rm H{\small II}{\it
phot}}' program will be published elsewhere
(\markcite{Thilker1999}Thilker {\it et al.} 1999).

\noindent{\it Luminosity Functions}

The observed \ha fluxes were converted into \ha luminosities using the
distances for NGC 1398 and NGC 7552 listed in Table 1. The striking
difference between the \ha images of NGC 1398 and NGC 7552 (see Figures
5 and 6 respectively) is also reflected in the widely dissimilar \hii
region luminosity functions as illustrated in Figure 8. In both cases, the {\it bright} 
\hii regions have a luminosity function that is well represented by a
power-law: $$ N(L) \propto L^{\alpha}dL $$ with exponent values of
-2.4($\pm 0.2$) and -2.0($\pm 0.1$) for NGC 1398 and NGC 7552
respectively. Our slope for NGC 1398 is
in good agreement with the slope obtained by
\markcite{Caldwell1991}Caldwell {\it et al.} for the same galaxy,
whereas our value is steeper for NGC 7552 than the -1.7 slope
determined by \markcite{Feinstein1997}Feinstein (1997).

The two categories of early-type spirals, as defined in this paper, are
based on the \ha luminosity of the largest \hii region in the
disk. The \hii region luminosity functions show that \hii regions,
in general, are more luminous in the category 2 galaxy, NGC 7552, as
compared to the category 1 galaxy, NGC 1398. We also, cautiously, report 
the difference in the shape of the two luminosity functions. 
NGC 1398 has a steep power law distribution. The \hii region luminosity function for 
NGC 7552, on the other hand, is a Gaussian with a turnover luminosity located at least 
a factor of six  higher than in NGC 1398. The two galaxies are at about 
the same distance and were observed on the same night with the 1.5m
telescope at CTIO for the same integration time. Similarly the data was 
reduced in an identical manner and \hii region luminosity functions
were determined with '{\rm H{\small II}{\it phot}}' using the same parameters
for the two galaxies. So there is little doubt that the the difference in 
the shapes of the two luminosity functions is real and confirms 
the visual impression given by \ha images of NGC 1398 and 
NGC 7552 in Figures 5 and 6 respectively.

Thus the difference in
global \ha luminosities between the two categories of early-type
spirals is not attributable to just one anomalous \hii region, but
rather suggests intrinsic differences in the ensemble of \hii regions
found in the two categories of early-type spirals. A subsequent paper
will present a more detailed analysis of the \hii region luminosity
functions for nearby early-type spirals.

\section{Discussion}

Early-type spirals have, in the past, been associated with low massive
star formation rates. The results presented in this paper, however,
indicate that early-type spirals are a heterogeneous group of galaxies with 
a significant number of them showing star forming regions that are 
comparable in luminosity to those found in late-type spirals. 
Early-type spirals have been divided into two categories based 
on the \ha luminosity of their largest \hii region in the disk. \hii regions 
in all category 1 galaxies have $L_{H\alpha} < 10^{39}\lum$, whereas 
category 2 early-type spirals have at least one \hii region with 
$L_{H\alpha} \ge 10^{39}\lum$. The purpose
of the following discussion is to elaborate on the diversity of early-type
spirals as revealed by the \ha and the continuum images.

\subsection{\it{Nuclear Diversity in Category 1 Galaxies }}

Most of the category 1 early-type spirals  appear to be morphologically
undisturbed. Despite the similarities in the continuum image, 
they have very diverse nuclear \ha properties ranging
from nuclear starbursts to low luminosity extended nuclear emission
line regions (ENERs) to unresolved point sources. A recent,
comprehensive, spectroscopic survey of the nuclei of nearby galaxies
by \markcite{Ho1997}Ho {\it et al.} (1997) revealed a similarly wide
variety of spectroscopic phenomena including \hii nuclei in $22\%$,
LINERs in $36\%$, and Seyfert nuclei in $18\%$ of early-type
spirals. In addition, they have found that LINERs and Seyferts reside
most frequently in early-type spiral galaxies. Unfortunately, we are
presently unable to correlate all of the images presented in this paper with
Ho's spectroscopic survey as most of our images are for the southern
hemisphere whereas Ho's survey is for the northern
hemisphere. However, we do have spectroscopic classifications 
for the nuclei of 17 galaxies that can be used to determine the correspondence 
between spectroscopy and morphology (Table 4).

All category 1 galaxies with unresolved \ha nuclear point sources have
been optically classified as Seyferts except NGC 5188, which
is classified as ``H'' by \markcite{VV1986}Veron Cetty \& Veron (1986). 
Spectroscopic information is available for only 4 ENER category
1 early-type spirals but all 4 have a LINER or a Seyfert-like
spectrum, classified as ``N'' by \markcite{VV1986}Veron-Cetty \& Veron
(1986).  The spectra of Seyfert-like, ``N'', galaxies have faint \ha
and [NII] lines, $H\alpha < 1.2 \times [NII] \lambda 6583$, and no
other detectable emission lines. The poor S/N of the data, make
it impossible to distinguish between a Seyfert 2 and a LINER
(\markcite{VV1986}Veron-Cetty
\& Veron 1986), however, we suspect that all ENER category 1 galaxies with
``N'' classification are LINERs. Conversely, we also suspect that not all
LINERs are AGNs, as advocated recently by \markcite{Ho1997}Ho {\it et al.} (1997).

The re-discovery of a possible correspondence between LINERs and
extended emission line gas in the nuclei of spiral galaxies is not
surprising. \markcite{Keel1983a}Keel (1983a), in an
\ha imaging survey of the nuclei of spiral galaxies hosting LINER spectra, 
found extended emission in most cases. In addition, he found that LINER type spectra
are ubiquitous in the nuclei of spiral galaxies not containing an
\hii region of much higher emission luminosity, which would make the
low ionization region undetectable if present
(\markcite{Keel1983b}Keel 1983b). Similarly, we have been able to
detect ENERs only in galaxies with few or no \hii regions in the
nuclear region.

M31, the nearest spiral galaxy, hosts an extended nuclear emission
line region(\markcite{RF1971}Rubin \& Ford 1971;
\markcite{Jacoby1985}Jacoby {\it et al.}  1985) and provides an
excellent opportunity to study the circumnuclear region in detail. The
ENER in M31 spans 2 kpc (\markcite{Devereux1994}Devereux {\it et al.}
1994) and has a complex filamentary structure that is similar to the
ENERs observed in more distant category 1 early-type spirals. The
spectroscopic analysis of the ENER in M31 shows that it is indeed a
LINER (\markcite{Heckman1996}Heckman 1996).  The absence of a UV
nuclear point source rules out an AGN as a source for photoionization
of the emission line gas in M31. Similarly, HST observations have
shown that the nuclear spiral in M31 is not ionized by massive stars
(\markcite{King1992}King {\it et al.} 1992;
\markcite{Devereux1994}Devereux {\it et al.} 1994).  Post-asymptotic
giant branch(PAGB) stars provide a possible source of ionization  for the
gas in the nuclear region (\markcite{Devereux1995}Devereux {\it et
al.} 1995; \markcite{Heckman1996}Heckman 1996) and
\markcite{Binette1994}Binette {\it et al.} (1994) have shown that
photoionization by PAGB stars can lead to a LINER spectrum. Thus a LINER 
spectrum does not necessarily betray an AGN as advocated recently by 
Ho {\it et al.} (1996).
 
M81 is the nearest category 1 Sa-Sab spiral and it hosts a low
luminosity LINER/Seyfert 1 nucleus (\markcite{FS1988}Filippenko \&
Sargent 1988; \markcite{Ho1996}Ho {\it et al.} 1996). The presence of
a broad line region(\markcite{PT1981}Peimbert \& Torres-Peimbert
1981), a compact X-ray (\markcite{Fabbiano1988}Fabbiano 1988;
\markcite{Petre1993}Petre {\it et al.} 1993) and radio 
source (\markcite{Beitenholz1996}Beitenholz {\it et al.}), and a UV nuclear source
(\markcite{Devereux1997}Devereux {\it et al.}  1997) strongly argue
for the presence of a low luminosity AGN (\markcite{Ho1996}Ho {\it et
al.} 1996; \markcite{Maoz1995}Maoz {\it et al.} 1995).  High
resolution \ha imaging has also revealed an extended nuclear emission
line spiral which is strikingly similar to the one observed in M31
(\markcite{Jacoby1989}Jacoby {\it et al.}  1989;
\markcite{Devereux1995}Devereux {\it et al.} 1995). The spiral is not
ionized by the spectroscopically identified AGN, as the
Seyfert contributes $< 5\%$ of the total \ha luminosity of the spiral
(\markcite{Devereux1995}Devereux {\it et al.} 1995;
\markcite{Devereux1997}Devereux {\it et al.}  1997). Furthermore, the
extended emission line gas, like M31, has a LINER spectrum(Heckman,
private communication). Recent HST observations
have shown that the nuclear spiral in M81 is not ionized by massive
stars (\markcite{Devereux1995}Devereux {\it et al.}  1995,
\markcite{Devereux1997}1997). The source of ionization for the nuclear
spiral is unknown, but it is most likely shock-ionized or photo-ionized
by UV radiation from bulge post asymptotic giant branch stars
(\markcite{Devereux1995}Devereux {\it et al.}  1995). M81 is likely to
be a composite object; a truly compact AGN surrounded by an extended
emission line region like the one in M31.

The two nearest LINERS, M31 and M81, illustrate the need for both
spectroscopic and imaging observations in order to distinguish
between true AGNs and ENERs with spectra that look like AGN.
The spectroscopic survey
conducted by \markcite{Ho1997a}Ho {\it et al.} (1997a) included most
of our northern hemisphere early-type spirals, which we are in the
process of imaging. The new images will allow us to determine the 
extent to which the spectroscopically identified AGNs
are responsible for ionizing the extended emission line gas seen under
the bulges of early-type spirals.

\subsection{\it{Morphological Peculiarities in Category 2 Galaxies}}

In recent years it has been suggested that interactions may play an important role
in the formation and subsequent evolution of early-type spirals (e.g.
\markcite{Schweizer1990}Schwiezer 1990, \markcite{Pfenniger1991}Pfenniger 1991, 
\markcite{Zaritsky1995}Zaritsky 1995). Furthermore, there is growing observational 
evidence  to suggest that
early-type spirals are perhaps the most dynamic of all the nearby
galaxy systems.  A recent HST survey of the bulges of nearby spiral
galaxies has revealed that a significant fraction ($\sim 40\%$) 
of early-type spirals show little or no morphological evidence for a
smooth, $R^{1/4}$ law (\markcite{Carollo1998}Carollo {\it et al.} 1998).  
Indeed, \markcite{SS1988}Schwiezer \& Seitzer (1988) discovered ripples 
around some early-type
spirals indicating that a major accretion event occurred in the past
1-2 Gyr. Similarly, the discovery of counter-rotating gas and star
disks in the early-type spirals NGC 4826 (\markcite{Braun1992}Braun {\it
et al.} 1992) and NGC 7217 (\markcite{MK1994}Merrifield \& Kuijken
1994) provide additional evidence for past interactions.

Perhaps galaxies classified as category 2 are the early-type spirals 
interacting in the current epoch. Note that we had excluded all galaxies that had
cataloged companions within 6\arcmin  and yet we still see signs of morphological
disturbance in most of the category 2 galaxies. The discovery of
previously unknown inter-loping dwarf galaxies in two of the category 2
early-type spirals provides direct observational evidence for on-going
interactions. Other morphological peculiarities,
such as tidal tails and dust lanes, are present in all category 2
galaxies providing further indirect evidence for interactions.
Numerous \ha and far-infrared studies indicate that interactions
elevate massive star formation
rates(e.g. \markcite{Bushouse1987}Bushouse 1987,
\markcite{Kennicutt1987}Kennicutt {\it et al.} 1987, \markcite{LK1995}Liu \& Kennicutt
1995, \markcite{Young1996}Young {\it et al.} 1996). All of the category 2 galaxies have
far-infrared luminosities in excess of $ 10^{10}L_{\sun}$, and
populate the high end of the local \ha luminosity function, both measures
indicative of high rates of massive star formation that is clearly
seen in \ha images in the form of giant \hii regions.
It is interesting to note that the distinction between the two 
categories of early-type spirals is based only on the luminosity of 
the largest \hii region in the disk, and yet this simple segregation 
leads to other clear differences in the range of global
\ha luminosities, galaxy morphology, and the range of 
individual \hii region luminosities.

The presence of morphological peculiarities in a significant 
fraction of early-type spirals may be linked to the idea of
forming big bulges through minor mergers. 
According to the numerical 
simulations of  \markcite{MH1994}Mihos
\& Hernquist (1994), the merger of a dwarf galaxy with a disk 
galaxy can lead to the formation of a stellar bulge. A series of such
mergers or accretion events could form the large bulges
found in  early-type spirals (\markcite{Pfenniger1993}Pfenniger 1993; 
\markcite{MH1994}Mihos \& Hernquist 1994; \markcite{RT1997}Rich \& Terndrup 1997).

Classical models of galaxy formation 
(\markcite{Eggen1962}Eggen {\it et al} 1962) predict bulges
of galaxies to be old and metal rich. While the Galactic Bulge provides
support for this theory, it may not be representative of all bulges. Recent observations
of bulges have revealed them to be diverse and heterogeneous in nature (for a review 
see \markcite{Wyse1997}Wyse {\it et al.} 1997).  Work on color gradients 
by \markcite{BP1994}Balcells \& Palatier (1994) show that there is little or no color change between 
the disk and bulge for some early-type spirals. Bulge populations seem to resemble 
their parent disks, suggesting that bulges may not be significantly older.
Similarly, blue bulges
observed in several S0 galaxies (\markcite{Schweizer1990}Schweizer 1990)
(NGC 5102, NGC 3156, IC 2035 etc.) 
lend further support to the idea of bulge formation through minor mergers. Further
work on stellar populations is needed to understand and quantify the diverse properties
exhibited by the bulges in nearby early-type spiral galaxies.

\subsection{{\it Influence of Bars on Early-type Spirals}}

Several studies have been conducted to determine the influence of bars
on star formation rates (e.g. \markcite{Hawarden1986}Hawarden {\it et al.} 1986; 
\markcite{Devereux1987}Devereux 1987; 
\markcite{Puxley1988}Puxley {\it et al} 1988; 
\markcite{Arsenault1989}Arsenault 1989; 
\markcite{RD1994}Ryder \& Dopita 1994; 
\markcite{Huang1996}Huang {\it et al.} 1996; 
\markcite{Tomita1996}Tomita {\it et al.} 1996; 
\markcite{Ho1997b}Ho {\it et al} 1997b ). Whereas, \ha and far-infrared studies 
of large galaxy samples (\markcite{RD1994}Ryder \& Dopita 1994; 
\markcite{Tomita1996}Tomita {\it et al.} 1996 respectively)
find no correlation of global star formation with the 
presence of bars, numerical models suggest that stellar bars
can strongly perturb gas flows in disk, and trigger nuclear star
formation. Indeed, nuclear star formation rates are measured
to be preferentially higher for galaxies that have bars, especially in
early-type spirals.  \markcite{Devereux1987}Devereux (1987) measured strong nuclear 
10$\mu m$ emission in 40$\%$ of barred early-type spirals, 
indicating high star formation rates. A similar excess was not
observed in late-type barred spirals. To provide some measure of
comparison the {\it nuclear} star formation rates in 
some barred early-type spirals are comparable to the globally 
integrated star formation rates of late-type spirals. 

Enhanced nuclear star formation in early-type spirals may be related
to the location of the  inner Lindblad resonance (ILR) 
(\markcite{Devereux1987}Devereux 1987) which is dependent
on the central mass distribution. The ILR in bulge-dominated Sa-Sab 
galaxies is expected to be located well inside the bar and
close to the nucleus (\markcite{EE1985}Elmegreen \& Elmegreen 1985), allowing 
a  stellar bar
to transfer gas from the disk region into the nuclear region more efficiently,
prompting a nuclear starburst. However, the presence of a  bar in an early-type spiral
does not necessarily mean enhanced nuclear star formation. NGC 1398 and
NGC 1433 (Figure 5) have prominent bars and yet they have little or no massive 
star formation in the nuclear region. Similarly, bars are not 
essential for the presence of a nuclear starburst as is illustrated by 
the starburst in the nucleus of non-barred galaxy, NGC 3885.

\ha observations of a large sample of early-type spirals provide an excellent 
opportunity to study the effects of bars on these bulge-dominated Sa-Sab 
galaxies. Unfortunately, with our current data,  we don't yet have enough statistics
to analyze the influence of bars with any certainty. In our sample
of 27 galaxies presented here, 17 have bars (including
intermediate cases), 5 don't have bars, and no information is available for 
the remaining 5 galaxies (see Table 4).
We will discuss early-type spirals in the context of bars in a later paper
when we have obtained \ha images for the complete sample of nearby 
early-type spirals. 

\subsection{{\it Star Formation in Early-type Spirals}}

The star forming properties of galaxies can also be quantified using \ha equivalent 
widths, where the \ha flux is normalized by the red light representing the 
older stellar population. In essence, the \ha equivalent width
measures the ratio of current star formation to past star formation. 
Figure 9 compares \ha equivalent widths for the galaxies
presented in this paper with the early-type spirals observed by
\markcite{KK1983}Kennicutt \& Kent (1983) and 
\markcite{Usui1998}Usui {\it et al.} (1998). Contrary to popular 
perception, \ha equivalent widths of early-type spirals presented 
in this paper reveal a significant fraction of galaxies with high ratios 
of present to past star formation. The comparison in Fig. 9
suggests that \markcite{KK1983}Kennicutt \& Kent's sample contains early-type spirals  
with preferentially low massive star formation rates. This has also been 
noted by \markcite{Usui1998}Usui {\it et al.} who have obtained 
\ha fluxes of galaxies with high L(fir)/L(blue) luminosity ratios and their 
results are also shown in Figure 9.  The images
presented in this paper reveal early-type spirals to encompass 
a wide range of \ha equivalent widths spanning that measured by both
Kennicutt \& Kent and Usui {\it et al}. 

In an attempt to reconcile the differences in \ha equivalent widths between 
this study and \markcite{KK1983}Kennicutt \& Kent (1983), we have examined 
the morphological classifications of the galaxies. 
We have noted that
\markcite{KK1983}Kennicutt \& Kent used classifications from the Revised Shapley-Ames 
Catalog (RSA) (\markcite{ST1981}Sandage \& Tammann 1981). On the other hand 
classifications in the NBG catalog (\markcite{Tully1988}Tully 1988), which are
used for this paper,  are derived
mostly from the Second Reference Catalog (RC2) (\markcite{devauc1976} de Vaucouleurs
{\it et al.} 1976). Table 5 lists the 
morphology of the galaxies included  in the present study as they appear in the  
Tully catalog (\markcite{Tully1988}Tully 1988), the Second Reference Catalog (RC2) 
(\markcite{devauc1976}de Vaucouleurs 1976),
and the Revised Shapley-Ames Catalog (\markcite{ST1981}Sandage \& Tammann 1981).
 Part of the difference in the 
\ha equivalent widths measured for early-type spirals by us and 
 \markcite{KK1983}Kennicutt \& Kent can be traced to a difference in galaxy classification.
Almost all of the category 2 Sa-Sab galaxies have been classified as Sb's in
the RSA catalog and would have been regarded as such in the study of 
\markcite{KK1983}Kennicutt \& Kent. 

The systematic difference between RC2 and RSA classifications is surprising
and may be traced to subtle variations in the application of the criteria used for 
classifying spiral galaxies. 
The classification criteria for the RC2 are explained in {\it Handbuch der Physik} 
(\markcite{devauc1959}de Vaucouleurs 1959), whereas RSA classifications are
elaborated in {\it the Hubble Atlas of Galaxies} (\markcite{Sandage1961}Sandage 1961).
In the RC2 catalog, sub-types of spirals, Sa-Sc, are defined by the ``{\it relative
importance of the nucleus (decreasing from a to c) and the degree
of unwinding and resolution of the arms (increasing from a to c)}''
(\markcite{devauc1959}de Vaucouleurs 1959). Spiral galaxies in RSA catalog are classified
according to the same criteria: ``{\it the openness of the spiral arms, the degree of resolution 
of the arms into stars, and the relative size of the unresolved nuclear 
region}'' (\markcite{Sandage1961}Sandage 1961). However, the third criterion (bulge size) was
not considered important for actual classifications in RSA; ``{\it For many 
years it was thought that the third classification criterion of the 
relative size of the unresolved nuclear region usually agreed 
with the criterion of the arms. Inspection of large numbers of 
photographs shows that, although there is a general correlation 
of the criteria, there are Sa galaxies that have small nuclear regions. 
The assignment of galaxies to the Sa, Sb, or Sc type is based here 
primarily on the characteristics of the arms}'' (Sandage 1961).
The preceding sentence may provide an explanation for the classification difference
found for category 2 galaxies in the RC2 and RSA catalogs. The differences can be
quite extreme, for example, NGC 7552 is considered 
a prototypical SBab galaxy by \markcite{devauc1959}de Vaucouleurs (1959) but 
an SBbc by Sandage in the RSA catalog(\markcite{ST1981}Sandage \& Tammann 1981).

Over the years, many studies have been conducted analyzing 
star forming properties of spiral galaxies along the Hubble 
sequence. \markcite{KK1983}Kennicutt \& Kent (1983), 
\markcite{dejong1984}de Jong (1984), \markcite{Sandage1986}Sandage (1986), 
\markcite{RL1986}Rieke \& Lebofsky (1986),
\markcite{PR1989}Pompea \& Rieke (1989), \& 
\markcite{Caldwell1991}Caldwell {\it et al.} (1991) used 
morphological classifications from RSA catalog and found
a dependence of star formation on Hubble types Sa-Scd.
However, this is not totally surprising since RSA classifications are 
primarily based on the arm characteristics, which are 
tightly correlated with star formation.
On the other hand,
\markcite{DY1990}Devereux and Young (1990), 
\markcite{IF1992}Isobe \& Fiegelson (1992),
\markcite{Tomita1996}Tomita {\it et al.} (1996), 
\markcite{Young1996}Young {\it et al.} (1996), Devereux and Hameed (1997), 
 and Usui {\it etal} (1998) have used RC2 classifications and find that 
star formation rates in spiral(Sa-Scd) galaxies are independent of Hubble type.
Similar results were obtained by   \markcite{Gavazzi1986}Gavazzi (1986) and 
\markcite{Bothun1989}Bothun {\it et al.}(1989),
who used classifications given in the UGC catalog. Thus it appears that 
perceptions concerning star formation rates among spiral galaxies 
along the Hubble sequence  depends on the choice of catalog. Furthermore, it may
be that other Hubble-type correlations also depend on the choice of catalog.

We do not want to claim that the classifications of one catalog are better 
than the other. However, we do want to stress the importance of the 
NBG catalog in the study of nearby galaxies. As described earlier, the NBG 
catalog is the most complete compilation of nearby galaxies. It contains
 all known galaxies within 40 Mpc that are brighter than 12th magnitude.
Any systematic study aimed at understanding the properties of 
nearby galaxies will rely on the NBG catalog. 

Results presented
in this paper have exposed the subjective nature of the prevailing
classification schemes. Perhaps it is time to start classifying 
galaxies quantitatively using measures of their physical properties.

\section{Summary \& Conclusions}

Preliminary results of an on-going \ha imaging survey of nearby,
bright, early-type (Sa-Sab) spirals have revealed them to be a
heterogeneous class of galaxies. Furthermore, our images have
identified a significant number of early-type spirals with massive star
formation rates comparable to the most prolifically star forming Sc
galaxies. An analysis of the \ha images suggests that early-type
spirals can be divided into two categories based on the \ha luminosity
of the largest \hii region in the disk. The first category includes
galaxies for which the individual \hii regions have \ha luminosity,
$L_{H\alpha} < 10^{39}
\lum$. Previously it was thought that all \hii regions in early-type spirals
have $L_{H\alpha} < 10^{39}\lum$, but our new observations have revealed 
a second category which includes $15-20\%$ of early-type spirals.

Most of the category 1 galaxies appear morphologically undisturbed. Despite
the similarities in the continuum images, category 1 galaxies exhibit 
diverse nuclear properties. The early-type spirals with the lowest 
\ha luminosities host extended nuclear emission line regions whereas the galaxies
that have the highest \ha luminosities contain nuclear starbursts. 
We also derived \hii region luminosity functions 
for one galaxy from each category. 
Our results suggest that, overall, \hii regions
in category 1 early-type spirals are less luminous than \hii regions
in category 2 early type spirals and that the difference between the 
two categories is not due to the existence of a single ``anomalous''
\hii region  with $L_{H\alpha} \ge 10^{39}\lum$. Dust lanes and tidal tails are 
present in the continuum images of all category 2 galaxies suggesting a 
recent interaction, which may also explain the presence of giant \hii
regions in these galaxies. 

The heterogeneous nature of early-type spirals, as revealed by 
the continuum and \ha images,  illustrates the problems associated
with classifying galaxies morphologically.  The results presented in this paper constitute
part of an on-going \ha survey to image a complete sample of nearby, bright, 
early-type spirals. It is anticipated that the survey will be the first 
to quantify the 
diverse nature of early-type spirals by providing a statistically 
significant set of observations.

\acknowledgments

The authors would like to thank Rene' Walterbos for providing the \ha
filter set for the APO observations. Thanks also to the TACs at NOAO and APO for the generous
allocation of observing time and their staffs for expert assistance at
the telescopes.  N.D. gratefully acknowledges the National Geographic
Society for travel support to CTIO.  S.H. would like to thank NOAO for
travel support to CTIO, Jon Holtzman, Charles Hoopes, and Bruce
Greenawalt for useful comments on the paper, and David Thilker for
providing the \hii region LF program, '{\rm H{\small II}{\it
phot}}'. We would also like to thank the referee,
Marcella Carollo, for useful comments that improved the
presentation of the paper. {\rm H{\small II}{\it phot}} was created by David Thilker
under support from NASA's Graduate Student Researcher Program
(NGT-51640).  The IDL source code and explanatory documentation will
soon be available by request.  Contact dthilker@nmsu.edu for details.

\newpage

\figcaption{Comparison of \ha flux measurements taken from the literature
with those measured from our \ha images using the same aperture size.  
NGC 1022 has two published values for its \ha flux (see text for details).}

\figcaption{Histograms illustrating the distribution of total \ha 
luminosity of Category 1 and Category 2 early-type spirals.}

\figcaption{Histograms illustrating the distribution of \ha equivalent widths
of Category 1 and Category 2 early-type spirals.}

\figcaption{Histograms illustrating the distribution of the nuclear(1kpc)
\ha luminosity of Category 1 and Category 2 early-type spirals.}

\figcaption{Red continuum and continuum-subtracted \ha images of 17 category 1
early-type spiral galaxies. The white bar at lower left in each image
represent 1 kpc in length. North is at the top and east is at the left in each image.}

\figcaption{ Red continuum and continuum-subtracted \ha images of 10 category 2 
early-type spiral galaxies. The white bar at lower left in each  image
represent 1 kpc in length. North is at the top and east is at the left in each image.}

\figcaption{Red continuum and continuum-subtracted \ha images of 2 'unclassified'
early-type spiral galaxies. The white bar at lower left in each image
represent 1 kpc in length. North is at the top and east is at the left in each image.}

\figcaption{The \hii region luminosity function of a Category 1 early-type spiral, NGC 1398, 
and a Category 2 early-type spiral, NGC 7552.}

\figcaption{A comparison of \ha equivalent widths of all early-type spirals
presented in this paper with the \ha equivalent widths of Kennicutt \& Kent (1983)
and Usui {\it et al.} (1998).} 

\begin{deluxetable}{lcccccc}
\scriptsize
\tablenum{1}
\tablewidth{0pt}
\tablecaption{Galaxy Parameters$^a$}
\tablehead{
\colhead{Galaxy} & \colhead{m(B)} & \colhead{size} & 
\colhead{$i$} & \colhead{V$_h$} & \colhead{Distance}\\
& \colhead{(mag)} & \colhead{(arc min)} &  & 
\colhead{($kms^{-1}$)} & \colhead{(Mpc)}
}
\startdata
NGC 660  &11.37    & 7.2 & 77$^\circ$ & 856 &  11.8  \nl
NGC 972  &11.75    & 3.9 & 65$^\circ$ & 1539 & 21.4 \nl
NGC 986  &11.66    & 3.3 & 42$^\circ$ & 1983 & 23.2 \nl
NGC 1022 &12.13    & 2.5 & 28$^\circ$ & 1503 & 18.5 \nl
NGC 1350 &11.16    & 5.0 & 62$^\circ$ & 1786 & 16.9 \nl
NGC 1371 &11.43    & 6.8 & 53$^\circ$ & 1472 & 17.1 \nl
NGC 1398 &10.47    & 7.6 & 50$^\circ$ & 1401 & 16.1 \nl
NGC 1433 &10.64    & 5.9 & 27$^\circ$ & 1071 & 11.6 \nl
NGC 1482 &13.50    & 2.1 & 58$^\circ$ & 1655 &  19.6\nl
NGC 1515 &11.17    & 5.7 & 89$^\circ$ & 1169 & 13.4 \nl
NGC 1617 &10.92    & 4.0 & 65$^\circ$ & 1040 & 13.4 \nl
NGC 2146 &11.00    & 5.3 & 36$^\circ$ & 918  &  17.2\nl
NGC 2273 &11.63    & 3.4 & 50$^\circ$ & 1844 &  28.4\nl
NGC 3169 &11.24    & 5.0 & 59$^\circ$ & 1229 &  19.7\nl
NGC 3471 &12.98    & 1.9 & 64$^\circ$ & 2076 &  33.0\nl
1108-48  &13.64    & 2.4 & 53$^\circ$ & 2717 &  35.2\nl
NGC 3717 &11.87    & 6.1 & 90$^\circ$ & 1731 &  24.6\nl
NGC 3885 &12.56    & 2.9 & 77$^\circ$ & 1948 &  27.8\nl
NGC 5156 &11.92    & 2.4 & 24$^\circ$ & 2983 &  39.5\nl
NGC 5188 &12.58    & 3.8 & 74$^\circ$ & 2366 &  32.9\nl
NGC 5728 &11.75    & 2.3 & 65$^\circ$ & 2970 &  42.2\nl
NGC 5915 &11.88    & 1.4 & 42$^\circ$ & 2272 &  33.7\nl
NGC 6810 &11.40    & 3.2 & 82$^\circ$ & 1995 &  25.3 \nl
NGC 7172 &12.55    & 2.1 & 64$^\circ$ & 2651 &  33.9 \nl
NGC 7213 &11.35    & 2.1 &  -   & 1778 & 22.0 \nl
NGC 7552 &11.31    & 3.5 & 31$^\circ$ & 1609 & 19.5 \nl
NGC 7582 &11.06    & 4.5 & 65$^\circ$ & 1459 & 17.6 \nl

\tablenotetext{a}{From Tully (1988)}

\enddata
\end{deluxetable}

\begin{deluxetable}{lcccccc}
\scriptsize
\tablenum{2}
\tablewidth{0pt}
\tablecaption{Details of Observations}
\tablehead{
\colhead{Galaxy}	& \colhead{Epoch}	& \colhead{Telescope} &
\colhead{H$\alpha$ filter} & \colhead{Exp. time} & \colhead{Standard} & 
\colhead{Ref.} 
}

\startdata
NGC 660  & 1996 Aug 14    & APO 3.5m & 6570/72 & 600s &  BD$+28^\circ$4211 & 1 \nl
NGC 972  & 1996 Aug 14    & APO 3.5m & 6610/70 & 600s &  BD$+28^\circ$4211 & 1\nl
NGC 986  & 1997 Oct 23    & CTIO 1.5m & 6606/75 & 900s & LTT 1020 & 2 \nl
NGC 1022 & 1997 Oct 25    & CTIO 1.5m & 6606/75 & 900s & LTT 7987 & 2 \nl
NGC 1350 & 1997 Oct 24    & CTIO 1.5m & 6606/75 & 900s & LTT 7987 & 2 \nl
NGC 1371 & 1997 Oct 25    & CTIO 1.5m & 6606/75 & 900s & LTT 7987 & 2 \nl
NGC 1398 & 1997 Oct 24    & CTIO 1.5m & 6606/75 & 900s & LTT 7987 & 2 \nl
NGC 1433 & 1997 Oct 26    & CTIO 1.5m & 6606/75 & 900s & LTT 7987 & 2 \nl
NGC 1482 & 1996 Nov 10    & APO 3.5m & 6610/70 & 720s &  G191B2B & 1 \nl
NGC 1515 & 1997 Oct 26    & CTIO 1.5m & 6606/75 & 900s & LTT 7987 & 2 \nl
NGC 1617 & 1997 Oct 23    & CTIO 1.5m & 6606/75 & 900s & LTT 1020 & 2 \nl
NGC 2146 & 1996 Nov 10    & APO 3.5m & 6570/72 & 720s &  G191B2B & 1 \nl
NGC 2273 & 1997 Jan 30    & APO 3.5m & 6610/70 & 550s &  G191B2B & 1 \nl
NGC 3169 & 1997 Mar 16    & CTIO 1.5m & 6606/75 & 900s & LTT 3218 & 2 \nl
NGC 3471 & 1997 Jan 30    & APO 3.5m & 6610/70 & 550s &  G191B2B & 1 \nl
1108-48  & 1997 Mar 15    & CTIO 1.5m & 6606/75 & 900s & LTT 3218 & 2 \nl
NGC 3717 & 1997 Mar 15    & CTIO 1.5m & 6606/75 & 900s & LTT 3218 & 2 \nl
NGC 3885 & 1997 Mar 16    & CTIO 1.5m & 6606/75 & 900s & LTT 3218 & 2 \nl
NGC 5156 & 1997 Mar 16    & CTIO 1.5m & 6606/75 & 900s & LTT 3218 & 2 \nl
NGC 5188 & 1997 Mar 15    & CTIO 1.5m & 6606/75 & 900s & LTT 3218 & 2 \nl
NGC 5728 & 1997 Mar 15    & CTIO 1.5m & 6649/76 & 900s & LTT 3218 & 2\nl
NGC 5915 & 1997 Mar 16    & CTIO 1.5m & 6606/75 & 900s & LTT 3218 & 2 \nl
NGC 6810 & 1997 Oct 24    & CTIO 1.5m & 6606/75 & 900s & LTT 7987 & 2 \nl
NGC 7172 & 1997 Oct 26    & CTIO 1.5m & 6606/75 & 900s & LTT 7987 & 2 \nl
NGC 7213 & 1997 Oct 25    & CTIO 1.5m & 6606/75 & 900s & LTT 7987 & 2 \nl
NGC 7552 & 1997 Oct 24    & CTIO 1.5m & 6606/75 & 900s & LTT 7987 & 2  \nl
NGC 7582 & 1997 Oct 23    & CTIO 1.5m & 6606/75 & 900s & LTT 7987 & 2  \nl

\tablerefs{
(1) Massey et al. 1988; (2) Hamuy et al. 1994.
}

\enddata
\end{deluxetable}

\begin{deluxetable}{lcccccc}
\scriptsize
\tablecolumns{7}
\tablenum{3}
\tablewidth{0pt}
\tablecaption{Summary of Results}
\tablehead{
\colhead{Galaxy}           &  \colhead{$F_{[H\alpha + N[II]]}^{a}$} & \colhead{Aperture}      
& \colhead{$L_{H\alpha}$}     & \colhead{$L_{H\alpha}(1kpc)$}    &
\colhead{$L_{H\alpha}(1kpc)/L_{H\alpha}$} \\
& \colhead{$(10^{-12}\flux)$} &  (arcsec)& \colhead{$(10^{40}\lum)$}& \colhead{$(10^{40}\lum)$} & 
}
\startdata
\cutinhead{Category 1 Early-type Spirals }
NGC 7213 & 2.9$\pm$0.4 &  210.7  & 16.8  & 5.2& $31\%$  \nl
NGC 3169 & 3.0$\pm$0.5 & 217.2 & 13.9 & 0.7 & $5\%$ \nl
NGC 5728 & 0.6$\pm$0.2 & 75.3 & 12.6 & 0.9 & $7\%$   \nl
NGC 3717 & 1.6$\pm$0.2 & 180.6 & 11.6 & 1.5 & $13\%$  \nl
NGC 5188 & 0.9$\pm$0.1 & 77.4 & 11.0 & 3.0 & $27\%$  \nl
NGC 1398 & 3.3$\pm$1.2 & 215.0 & 10.3 & 0.3 & $3\%$ \nl
NGC 3471 & 0.7$\pm$0.1  & 20.4  & 8.9   & 5.1 & $57\%$ \nl
NGC 3885 & 0.9$\pm$0.1 & 51.6 & 8.5 & 5.7 & $67\%$  \nl
NGC 1482 & 1.6$\pm$0.2 &  48.9  & 7.4 & 3.7 & $50\%$  \nl
NGC 1022 & 1.2$\pm$0.2 & 77.4 & 4.9 & 4.0 & $82\%$   \nl
NGC 1371 & 1.4$\pm$0.5 & 193.5 & 4.9 & 0.3 & $6\%$ &  \nl
NGC 1350 & 1.3$\pm$0.7 & 176.3 & 4.5 & 0.2 & $4\%$ \nl
NGC 1433 & 2.3$\pm$0.8 & 197.8 & 3.7 & 0.7 & $19\%$  \nl
NGC 7172 & 0.3$\pm$0.1 & 53.8 & 3.6 & 0.7 & $19\%$ \nl
NGC 2273 & 0.7$\pm$0.2 & 103.7 & 2.3 & 1.6 & $70\%$  \nl
NGC 1515 & 0.9$\pm$0.2 & 307 $\times$ 172 & 1.8 & 0.4 & $22\%$ \nl
NGC 1617 & 0.6$\pm$0.6 & 111.8 & 1.4 & 0.2 & $14\%$  \nl
\cutinhead{Category 2 Early-type Spirals }
NGC 5915 & 3.3$\pm$0.2 & 43.0 & 44.9 & 2.7 & $6\%$ \nl
NGC 5156 & 2.2$\pm$0.1 & 103.2 & 41.2 & 0.3 & $0.7\%$  \nl
NGC 7552 & 7.0$\pm$0.7 & 116.1 & 31.9 & 16.0 & $50\%$ \nl
NGC 986 & 3.3$\pm$0.4 & 114.0 & 21.3 & 6.3 & $30\%$ \nl
NGC 972 & 3.7$\pm$0.2 &  58.6  & 20.3 & 1.8 & $9\%$  \nl
NGC 6810 & 2.6$\pm$0.2 & 86.0 & 20.0 & 6.9 & $35\%$ \nl
1108-48 & 1.2$\pm$0.1 & 137.6 & 17.8 & 0.6 & $3\%$ \nl
NGC 7582 & 4.6$\pm$0.7 & 139.8 & 17.1 & 7.4 & $43\%$  \nl
\cutinhead{Unclassified${^b}$}
NGC 2146 & 4.6$\pm$1.0 & 119.0  & 16.3 & 2.8 & $17\%$ \nl
NGC 660 & 2.2$\pm$0.3 & 163 $\times$ 253 & 3.7 & 0.6 & $16\%$ \nl
\enddata
\tablenotetext
{a}{Fluxes have not been corrected for Galactic or internal extinction.} 
\tablenotetext
{b}{These galaxies exhibit characteristics of both categories. See text for
details.}
\end{deluxetable}

\begin{deluxetable}{lccccl}
\scriptsize
\tablecolumns{6}
\tablenum{4}
\tablewidth{0pt}
\tablecaption{Summary of Results II}
\tablehead{
\colhead{Galaxy}           &  \colhead{Nuc. morph.${^a}$}      &
\colhead{Nuc. Sp. class}          & \colhead{Ref.}  &
\colhead{Bar}        &  \colhead{Comments} 
}
\startdata
\cutinhead{Category 1 Early-type Spirals }
NGC 7213 & PS/ENER &  S1 & 1 & N & A giant \ha filament is located approximately 17 kpc 
south of the galaxy \\
& & & & & with no counterpart in the continuum image.\nl
NGC 3169 & ENER & L2 & 2 & N  & \nl
NGC 5728 & PS & S2 & 3 & X &  A tightly wound inner spiral arm and a loose outer spiral
arm in the continuum.\nl
NGC 3717 & ? & H & 4 & N &  A prominent dust lane parallel to the major axis of the galaxy\nl
NGC 5188 & PS & H & 4 & Y &  \nl
NGC 1398 & ENER & N & 5 & Y &  A bar-ring morphology in \ha.\nl
NGC 3471 & SB & H  & 6 & ? &  \nl
NGC 3885 & SB & H & 7 & N & \nl
NGC 1482 & SB & \nodata & \nodata & ? & A prominent dust lane parallel to the major axis of the galaxy. \\
& & & & & Filaments and/or chimneys of ionized gas extending perpend. to the disk.\nl
NGC 1022 & SB & H & 8 & Y & A tightly wound inner spiral arm and a loose outer spiral
arm in the continuum.\nl
NGC 1371 & ENER & \nodata & \nodata & X & \nl
NGC 1350 & ENER & N & 4 & Y &  A bar-ring morphology in \ha.\\
& & & & &  A tightly wound inner spiral arm and a loose outer spiral arm in the continuum.\nl
NGC 1433 & ENER & N & 4 & Y & A bar-ring morphology in \ha. \\
& & & & & A tightly wound inner spiral arm and a loose outer spiral
arm in the continuum.\nl
NGC 7172 & PS & S2 & 8 & ? & A prominent dust lane parallel to the major axis of the galaxy \nl
NGC 2273 & PS & S2 & 2 & Y & \nl
NGC 1515 & ENER & \nodata & \nodata & X & \nl
NGC 1617 & ENER & \nodata & \nodata & ? & \nl
\cutinhead{Category 2 Early-type Spirals }
NGC 5915 & \nodata & \nodata & \nodata & Y & Asymmetric spiral arms in the continuum. \\
& & & & & \ha morphology does not correspond with the major continuum features. \nl
NGC 5156 & \nodata & \nodata & \nodata & Y &  \nl
NGC 7552 & \nodata & H & 4 & Y & A dwarf galaxy at the end of the northern spiral arm \nl
NGC 986 & \nodata & H & 4 & Y &  A tidally disrupted dwarf galaxy at the end of the northern arm?\nl
NGC 972 & \nodata & H & 2 & ? & Dusty morphology. A possible bar-like structure crossing the nucleus \nl
NGC 6810 & \nodata & S2 & 4 & N & A prominent dust lane parallel to the major axis of the galaxy \nl
1108-48 & \nodata & \nodata & \nodata & Y & A faint tidal tail leading to a star forming region\\
&&&&& 18 kpc north-east of the nuc. \nl
NGC 7582 & \nodata & S2 & 9 & Y & A prominent dust lane parallel to the major axis of the galaxy \nl
\cutinhead{Unclassified${^b}$}
NGC 2146 & \nodata & H & 2 & Y & Highly disturbed morphology with a prominent dust lane \nl
NGC 660 & \nodata & T2/H & 2 & Y &  Highly disturbed morph.Two prominent dust lanes perpend. to 
each other\nl
\enddata

\tablenotetext
{a}{Nuclear regions of all category 2 and unclassified galaxies are resolved in \ha, 
but they do not meet the starburst or the ENER criteria \\
as defined in the text. Hence 
their nuclear regions have not been classified morphologically.}
\tablenotetext 
{b}{These galaxies exhibit characteristics of both categories. See text for
details.}

\tablecomments{ \ha nuclear morphology: PS=Unresolved point source, ENER=Extended
Nuclear Emission Line Region, SB=nuclear starburst\\ \\
Classification of the nuclear spectrum: S=Seyfert,
H=HII nucleus, L=LINER, T=Transition, N=Seyfert-like, i.e. H$\alpha < 
1.2 \times [NII] \lambda 6583$\\ \\
Bar classifications are taken from Tully (1988) : Y=Bar is present, N=Bar is absent,
X=intermediate case, ?=No information.}
\tablerefs{
(1)\markcite{Filippenko1984}Filippenko \& Halpern 1984; (2)\markcite{Ho1997a} Ho, Filippenko, and Sargent 1997;
(3) \markcite{Phillips1983}Phillips, Charles, and Baldwin 1983;\\
(4)\markcite{VV1986}Veron-Cetty \& Veron 1986; (5)\markcite{Balzano1983}
Balzano 1983; (6) \markcite{Lehnert1995}Lehnert \& Heckman 1995; (7) \markcite{Ashby1995}Ashby et al. 1995;
(8) \markcite{Sharples1984}Sharples, \\Longore, and Hawarden 1984; (9) \markcite{Unger1987}Unger et al. 1987 
}

\end{deluxetable}

\begin{deluxetable}{lcccc}
\scriptsize
\tablenum{5}
\tablewidth{0pt}
\tablecaption{Morphological Classifications}
\tablehead{
\colhead{Galaxy}           &  \colhead{NBG$^{a}$} & \colhead{RC2$^{b}$}      
& \colhead{RSA$^{c}$} 
 }
\startdata
\cutinhead{Category 1 Early-type Spirals }
NGC 7213 & Sa & Sa & Sa \nl
NGC 3169 & Sa & Sa & Sb  \nl
NGC 5728 & Sa & Sa & Sb  \nl
NGC 3717 & Sab & Sb & Sb  \nl
NGC 5188 & Sa & Sb & Sbc  \nl
NGC 1398 & Sab & Sab & Sab  \nl
NGC 3471 & Sa & Sa & \nodata  \nl
NGC 3885 & Sa & S0/a & Sa \nl
NGC 1482 & Sa & S0/a & \nodata  \nl
NGC 1022 & Sa & Sa & Sa  \nl
NGC 1371 & Sa & Sa & Sa  \nl
NGC 1350 & Sab & Sab & Sa \nl
NGC 1433 & Sa & Sa & Sb \nl
NGC 7172 & Sab & Sab & \nodata \nl
NGC 2273 & Sa & S0/a & \nodata  \nl
NGC 1515 & Sa & Sbc & Sb \nl
NGC 1617 & Sa & Sa & Sa  \nl
\cutinhead{Category 2 Early-type Spirals }
NGC 5915 & Sab & Sab & Sbc \nl
NGC 5156 & Sa & Sab & Sbc  \nl
NGC 7552 & Sab & Sab & Sbc\nl
NGC 986 & Sab & Sab & Sb \nl
NGC 972 & Sab & S0/a & Sb  \nl
NGC 6810 & Sa & Sab & Sb\nl
1108-48 & Sab & \nodata & \nodata  \nl
NGC 7582 & Sab & Sab & Sab \nl
\cutinhead{Unclassified}
NGC 2146 & Sab & Sab & Sb \nl
NGC 660 & Sa & Sa & \nodata  \nl
\enddata
\tablenotetext
{a}{{\it Nearby Galaxies Catalog}, Tully 1988.} 
\tablenotetext
{b}{{\it Second Reference Catalogue of Bright Galaxies}, de Voucouleurs 1976.}
\tablenotetext
{c}{{\it A Revised Shapley-Ames Catalog of Bright Galaxies}, Sandage \& Tammann 1981.}
\end{deluxetable}

\begin{figure}
\plotone{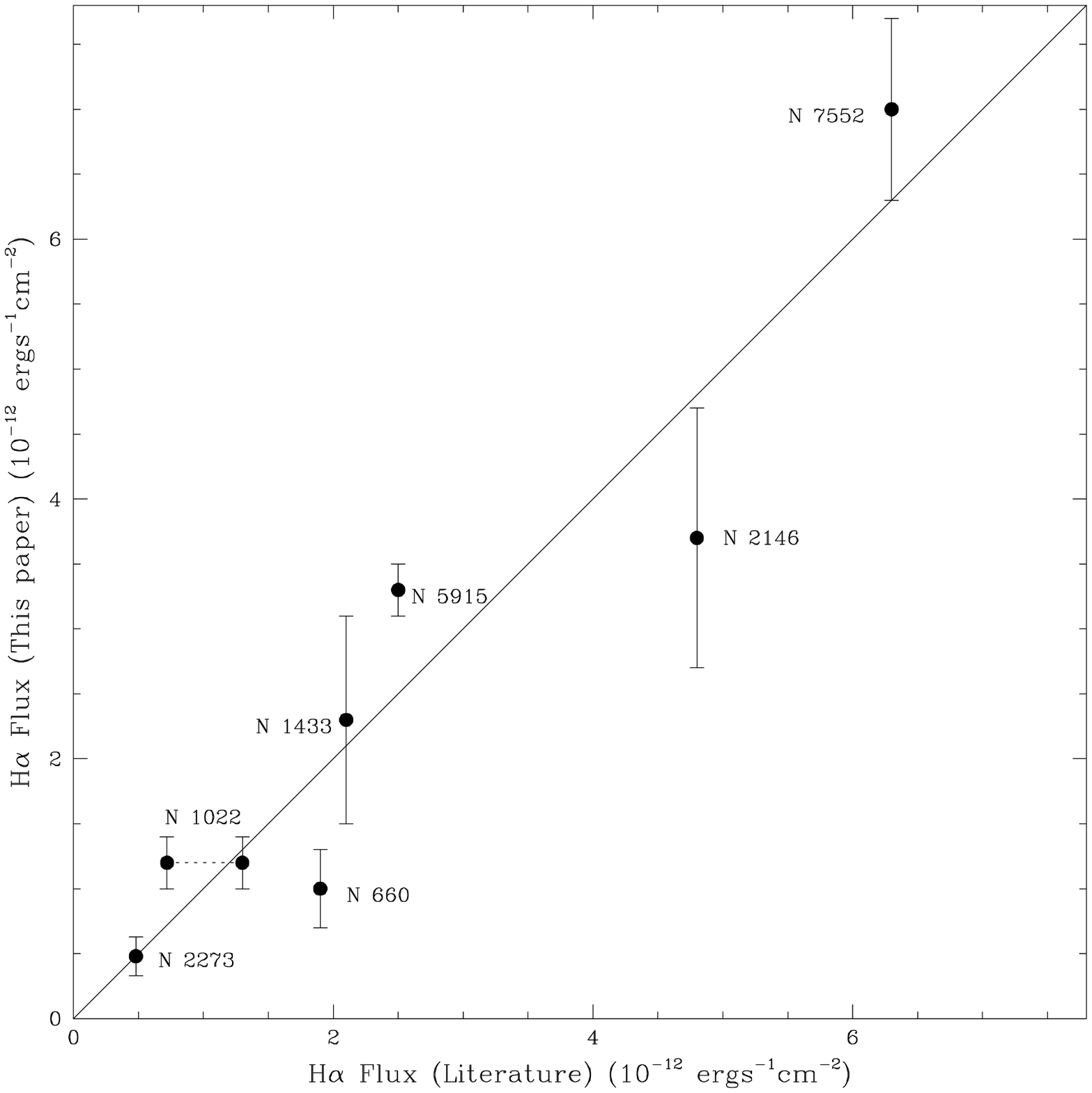}
\end{figure}

\begin{figure}
\plotone{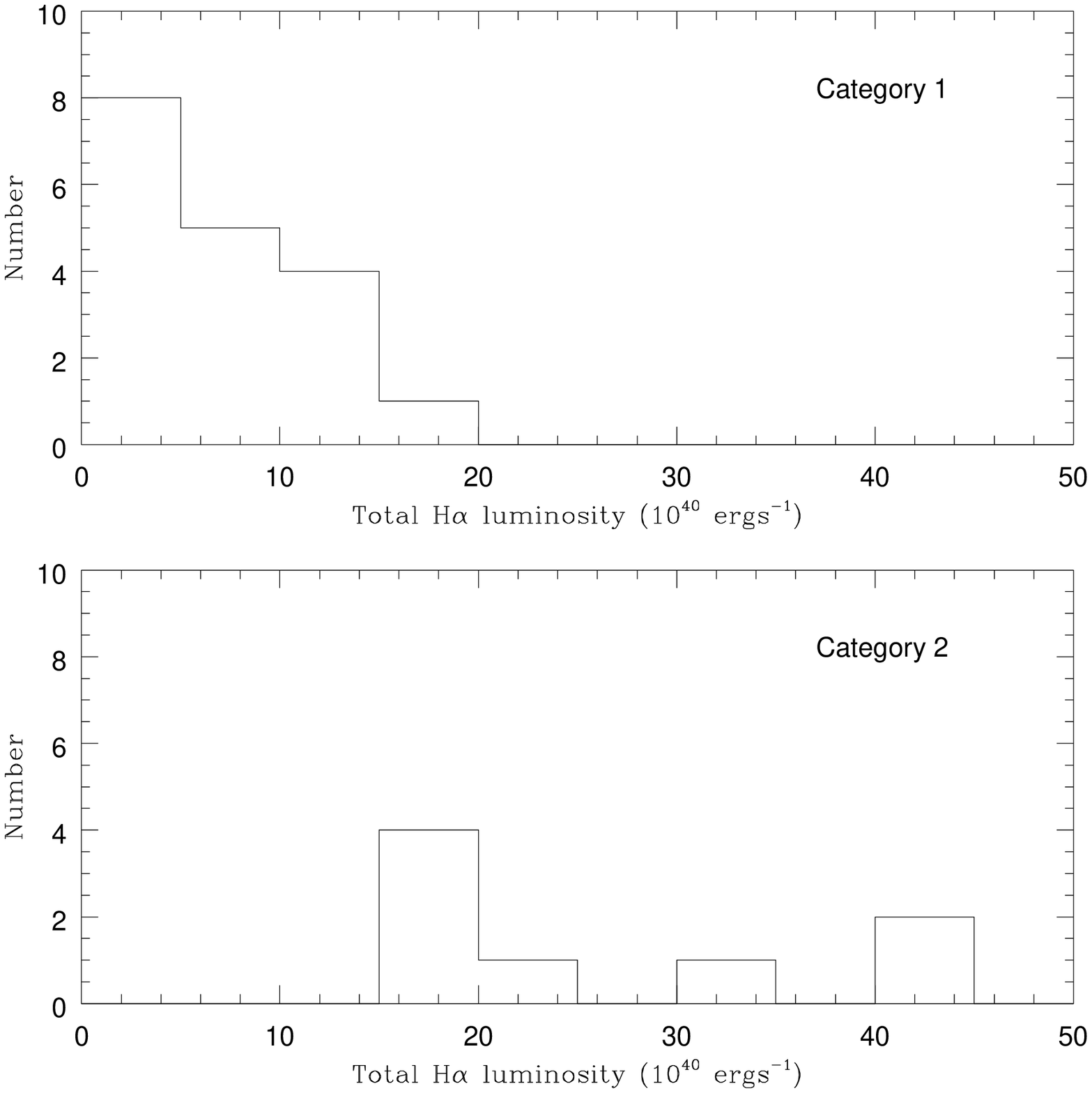}
\end{figure}

\begin{figure}
\plotone{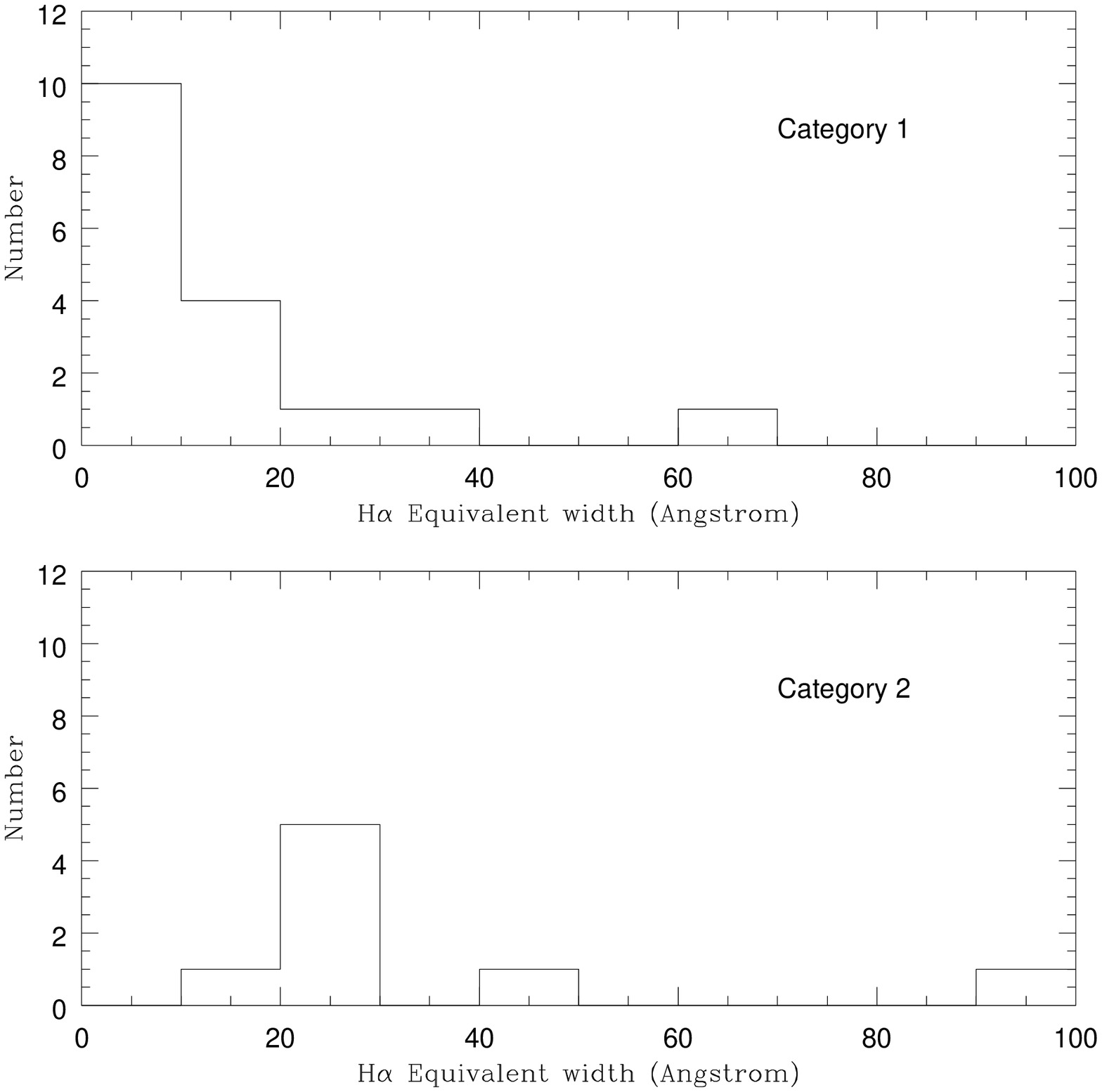}
\end{figure}

\begin{figure}
\plotone{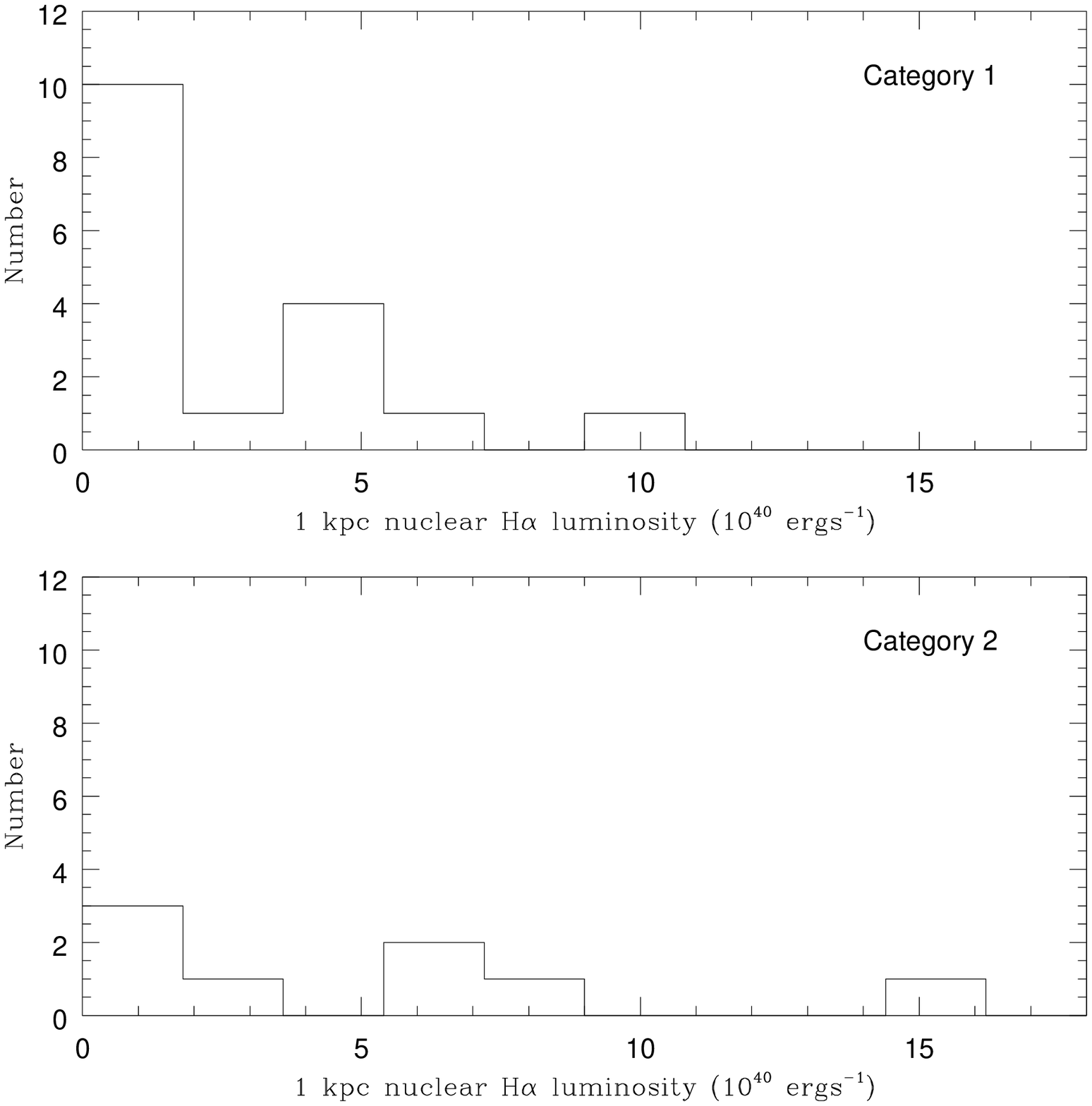}
\end{figure}

\begin{figure}
\plotone{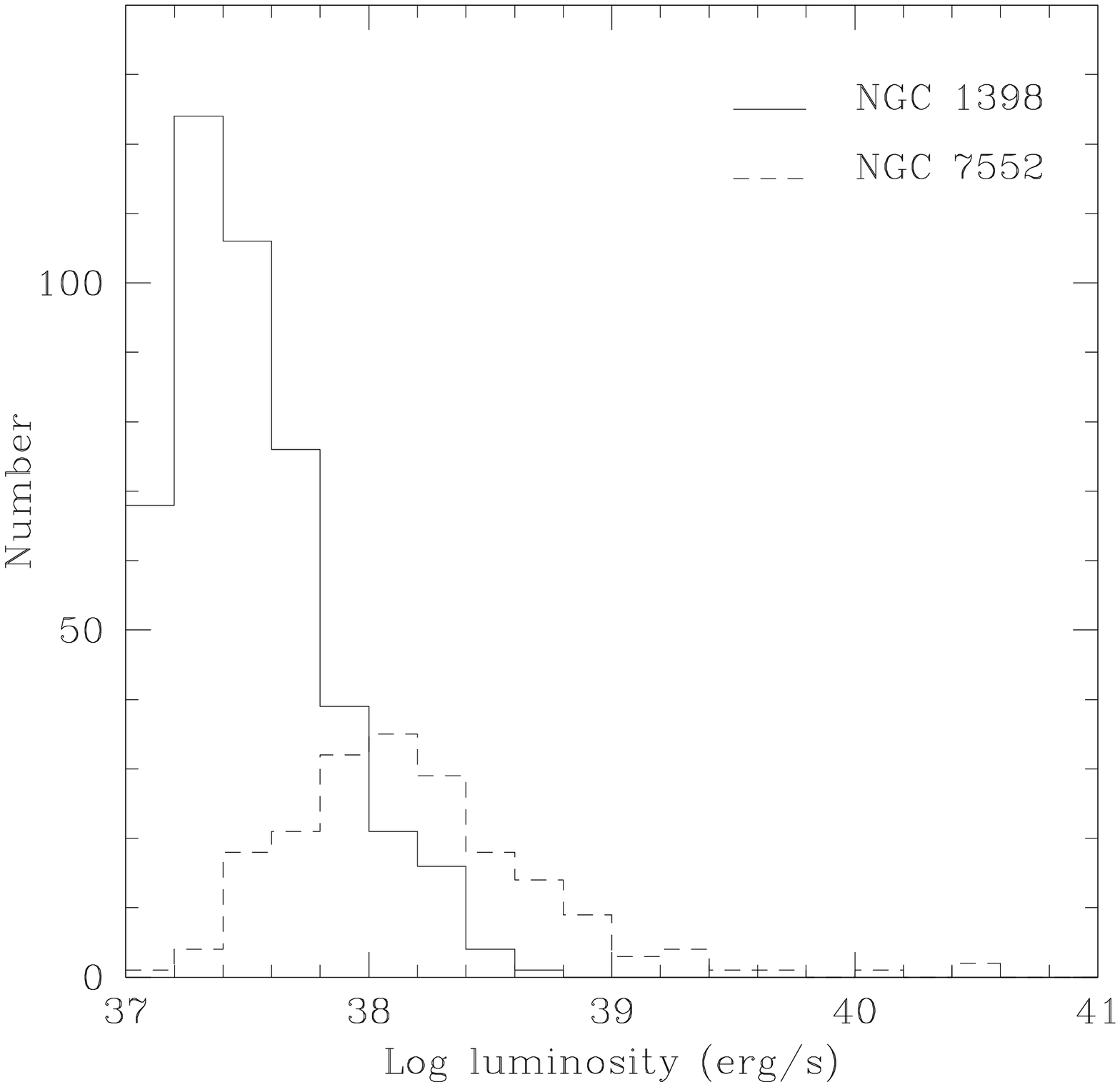}
\end{figure}

\begin{figure}
\plotone{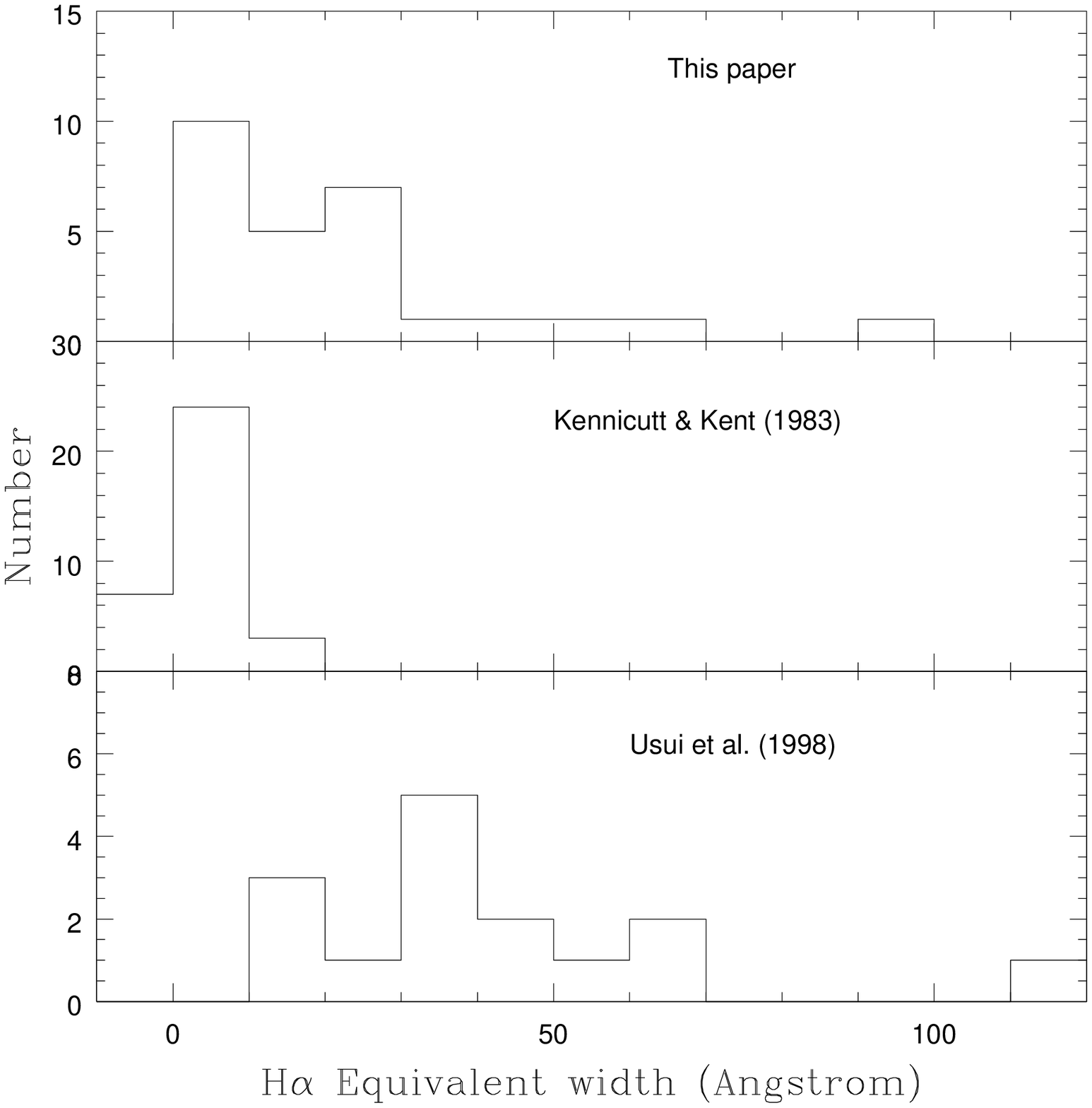}
\end{figure}


\begin{references}
\reference{Arsenault1989}Arsenault,R. 1989, A\&A, 217, 66
\reference{Ashby1995} Ashby,M.L.N., Houck,J.R., \& Matthews,K. 1995, \apj, 447, 545
\reference{BP1994} Balcells,M., \& Peletier,R. 1994, \aj, 107, 135
\reference{Balzano1983} Balzano,V.A. 1983, \apj, 268, 602
\reference{Beitenholz1996}Beitenholz,M.F., et al. 1996, \apj, 457, 604
\reference{Binette1994} Binette,L., Magris,C.G., Stasinska,G., \& Bruzual,A.G. 1994, A\&A, 292, 13
\reference{Bothun1989}Bothun,G.D., Lonsdale,C.J., \& Rice,W. 1989, \apj, 341, 129
\reference{Braun1992} Braun,R., Walterbos,R.A.M., \& Kennicutt,R.C.,Jr. 1992, 
Nature, 360, 442
\reference{Bresolin1997} Bresolin,F., \& Kennicutt,R.C.,Jr. 1997, \aj 113, 975
\reference{Bushouse1987} Bushouse,H.A. 1987, \apj, 320, 49
\reference{Caldwell1991} Caldwell,N, Kennicutt,R.C.,Jr, Phillips,A.C., \& 
Schommer,R.A. 1991, \apj, 370, 526 
\reference{Carollo1998} Carollo,C.M., Siavelli,M, \& Mack,J. 1998, \aj, 116, 68
\reference{Casoli1998} Casoli,F., Sauty,S., Gerin,M., Boselli,A., Fouque,P., Braine,J.,
Gavazzi,G., Lequeux,J., \& Dickey,J. 1998, A\&A, 331, 451
\reference{Crocker1996} Crocker,D.A., Baugus,P.D., \& Buta,R. 1996, \apjs, 105, 353
\reference{Devereux1987}Devereux,N.A. 1987, \apj, 323, 91
\reference{Devereux1997} Devereux,N.A., Ford,H., \& Jacoby,G. 1997, \apj, 481, L71
\reference{DH1997} Devereux,N.A., \& Hameed,S. 1997, \aj, 113, 599
\reference{Devereux1995} Devereux,N.A., Jacoby,G., \& Ciardullo,R. 1995, \aj, 110, 1115
\reference{Devereux1994} Devereux,N.A., Price,R., Wells,L.A., \&  Duric,N. 1994, \aj, 108, 1667
\reference{DY1990}Devereux,N.A., \& Young,J.S. 1990, \apj, 350, L25
\reference{Eggen1962} Eggen,O., Lynden-Bell,D., \& Sandage,A., 1962, \apj, 136, 748
\reference{EE1985} Elmegreen,B.G., \& Elmegreen,D.M. 1985, \apj, 288, 438
\reference{Fabbiano1988} Fabbiano,G., 1988, \apj, 325, 544
\reference{Feinstein1997} Feinstein,C., 1997, \apjs, 112, 29
\reference{Filippenko1984} Filippenko,A.V., \& Halpern,J.P. 1984, \apj, 285,458
\reference{FS1988} Filippenko,A.V., \& Sargent,W.L.W. 1988, \apj, 324, 134
\reference{Gavazzi1986}Gavazzi,G., Cocito,A., \& Vettolani,G. 1986, \apj, 305, L15
\reference{Hamuy1994} Hamuy,M., Suntzeff,N.B., Heathcote,S.R., Walker,A.R., Gigoux,P., 
\& Phillips,M.M. \pasp, 106, 566
\reference{Hawarden1986}Hawarden,T.G., Mountain,C.M., Leggett,S.K., \& Puxley,P.J. 1986,
\mnras, 221, 41
\reference{Heckman1996} Heckman,T.M. 1996, in {\it The Physics of LINERS in view of Recent Observations}, ed. Eracleous,M., Koratkar,A., Leitherer,C., \& Ho,L. (ASP Conference Series Volume 103)
\reference{Ho1996} Ho,L.C., Filippenko,A.V., \& Sargent,W.L.W. 1996, \apj, 462, 183 
\reference{Ho1997} Ho,L.C., Filippenko,A.V., \& Sargent,W.L.W. 1997, \apj, 487, 568
\reference{Ho1997a} Ho,L.C., Filippenko,A.V., \& Sargent,W.L.W. 1997a, \apjs, 112, 315
\reference{Ho1997b}Ho,L.C., Filippenko,A.V., \& Sargent,W.L.W. 1997b, \apj, 487, 591
 \reference{HK1983} Hodge,P., \& Kennicutt,R.C.,Jr. 1983, \aj, 88, 296
\reference{Huang1996}Huang,J.H., Gu,Q.S., Su,H.J., Hawarden,T.G., Liao,X.H., \&
Wu,G.X. 1996, A\&A, 313, 13
\reference{Hubble1936} Hubble, E. 1936, {\it The Realm of the Nebulae} (New 
Haven: Yale Univ. Press)
\reference{IF1992}Isobe,T., \& Feigelson,E.D. 1992, \apjs, 79, 197
\reference{Jacoby1989} Jacoby,G.H., Ciardullo,R., Ford,H.C., \& Booth,J. 1989, \apj, 344, 704
\reference{Jacoby1985} Jacoby,G.H., Ford,H.C., \& Ciardullo,R. 1985, \apj, 290, 136
\reference{dejong1984} de Jong,T. {\it et al.} 1984, \apj, 278, 67L
\reference{Keel1983a} Keel,W.B. 1983a, \apj, 268, 632
\reference{Keel1983b} Keel,W.B. 1983b, \apjs, 52, 229
\reference{Kennicutt1983} Kennicutt,R.C.,Jr. 1983, \apj, 272, 54
\reference{Kennicutt1988} Kennicutt,R.C.,Jr. 1988, \apj, 334, 144
\reference{Kennicutt1998} Kennicutt,R.C.,Jr. 1998, {\it Ann. Rev.
Astron. Astrophys.} in press.
\reference{Kennicutt1989} Kennicutt,R.C.,Jr., Edgar,K.B., \& Hodge,P.W., 1989, 337, 761
\reference{Kennicutt1987} Kennicutt,R.C.,Jr., Keel,W.C., van der Hulst,J.M., 
Hummel.E., \& Roettiger,K.A. 1987, \aj, 93, 1011
\reference{KK1983} Kennicutt,R.C.,Jr., \& Kent,S.M. 1983, \aj, 88, 1094
\reference{Kennicutt1994} Kennicutt,R.C.,Jr., Tamblyn,P., \& Congdon.C.W. 1994,
\apj, 435, 22 
\reference{King1992} King,I.R., et al. 1992, \apj, 397, L35
\reference{Kormendy1977} Kormendy,J. 1977, in {\it The Evolution of Galaxies and
Stellar Populations}, ed. B.M. Tinsley \& R.B. Larson (New Haven: Yale
Univ. Obs.), 131
\reference{Larson1974} Larson,R.B., \& Tinsley,B.M. 1974, \apj, 192, 293
\reference{Lehnert1995} Lehnert,M.D., \& Heckman,T.M. 1995, \apjs, 97, 89
\reference{LK1995} Liu,C.T., \& Kennicutt,R.C.,Jr. 1995, \apj, 450, 547
\reference{Maoz1995} Maoz,D., et al. 1995, \apj, 440, 91
\reference{Massey1988} Massey,P., Strobel,K., Barnes,J.V., \& Anderson,E. 1988,
\apj, 328, 315
\reference{MK1994} Merrifield,M.R., \& Kuijken,K. 1994, \apj, 432, 575
\reference{MH1994} Mihos,J.C., \& Hernquist,L. 1994, \apj, 425, L13
\reference{PT1981} Peimbert,M., \& Torres-Peimbert,S. 1981, \apj, 245, 845
\reference{Petre1993} Petre,R., Mushotzky,R., Serlemitsos,P.J., Jahoda,K., \& 
Marshall,F.E. 1993, \apj, 418, 644
\reference{Pfenniger1991} Pfenniger, D. 1991, in {\it Dynamics of Disc Galaxies}, 
ed. B. Sundelius, Goteborgs, p. 191
\reference{Pfenniger1993} Pfenniger, D. 1993, in IAU Symp. 153, {\it Galactic 
Bulges}, eds. H. Dejonghe and H.T. Habing (Dordrecht, Kluwer), p. 387
\reference{Phillips1983} Phillips,M.M., Charles,P.A., \& Baldwin,J.A. 1983, \apj, 
266, 485
\reference{PR1989}Pompea,S.M., \& Rieke,G.H. 1989, \apj, 342, 250
\reference{Puxley1988}Puxley,P.J., Hawarden,T.G., \& Mountain,C.M. 1988, \mnras, 
231, 465
\reference{RT1997} Rich,M.R., \& Terndrup,D.M. 1997, \pasp, 109, 571
\reference{Rieke1980} Rieke,G.H., \& Lebofsky,M.J. 1986, \apj, 304, 326
\reference{Rieke1980} Rieke,G.H., Lebofsky,M.J., Thompson,R.I., Low,F.J., \& Tokunaga,A.T.
1980, \apj, 238, 24
\reference{Roberts1969} Roberts,M.S. 1969, \aj, 74, 859
\reference{Roberts1994} Roberts,M.S., \& Haynes,M.P. 1994, {\it Ann. Rev.
Astron. Astrophys.}, 32, 115
\reference{RF1971} Rubin,V.C., \& Ford,H. 1971, \apj, 170, 25
\reference{RD1994}Ryder,S.D., \& Dopita,M.A. 1994, \apj, 430, 142
\reference{Sandage1961}Sandage,A.R. 1961, {\it The Hubble Atlas of Galaxies}, Publ. No. 618
(Carnegie Institution of Washington, Washington,D.C.)
\reference{Sandage1986}Sandage,A.R., A\&A, 161, 89
\reference{ST1981}Sandage,A., \& Tammann,G.A. 1981, {\it A Revised Shapley-Ames Catalog of
Bright Galaxies}, Publ. No. 635(Carnegie Institution of Washington, Washington,D.C.) (RSA)
\reference{Schweizer1978} Schweizer,F., 1978, \apj, 220, 98
\reference{Schweizer1990} Schweizer,F. 1990, in {\it Dynamics and Interactions
of Galaxies}, ed. Wielen.R. (Heidelberg: Springer-Verlag)
\reference{SS1988} Schweizer,F., \& Seitzer,P. 1988, \apj, 328,88
\reference{Sharples1984} Sharples,R.M., Longmore,A.J., Hawarden,T.G., and Carter,D.
1984, \mnras, 208, 15
\reference{Thilker1999} Thilker,D., Braun,R., Walterbos,R.A.M., \& Fierro,V. 1999, {\it in prep.}
\reference{Tomita1996} Tomita, A., Tomita, Y., and Saito, M. 1996, \pasj, 48, 285
\reference{Tully1988} Tully,R.B. 1988, {\it Nearby Galaxies Catalog} (Cambridge:
Cambridge University Press)
\reference{Unger1987} Unger,S.W., Lawrence,A., Wilson,A.S., Elvis,M., and Wright,A.E.
1987, \mnras, 228, 521
\reference{Usui1998} Usui, T., Saito, M., and Tomita, A. 1998, \aj, 116, 2166
\reference{devauc1959} de Vaucouleurs, G. 1959, {\it Handbuch der Physik}, 
vol. 53, 275 (Berlin: Springer-Verlag)
\reference{devauc1976}  de Vaucouleurs, G.,  de Vaucouleurs,A., \& Corwin,H.G. 1976,
{\it Second Reference Catalog of Bright Galaxies} (University of Texas, Austin) (RC2)
\reference{vendenbergh1976} van den Bergh,S. 1976, \aj, 81, 797
\reference{VV1986} Veron-Cetty,M.P., \& Veron,P. 1986, A\&AS, 66,335
\reference{Wyse1997} Wyse,R.F.G., Gilmore,G., \& Franx,M. 1997, {\it Ann. Rev.
Astron. Astrophys.}, 35, 637
\reference{Young1996} Young,J.S., Allen,L., Kenney,J.D.P., Lesser,A., \& Rowand,B.
1996, \aj, 112, 1903
\reference{Young1988} Young,J.S., Kleinmann,S.G., \& Allen.L.E. 1988, \apj, 334, L63
\reference{YK1989} Young,J.S., \& Knezek,P.M. 1989, \apj, 347, L55
\reference{Zaritsky1995} Zaritsky, D. 1995, \apj, 448, L17

\end{references}
\end{document}